\documentclass[aps,prx,twocolumn,superscriptaddress]{revtex4-2}

\usepackage{amsmath}
\usepackage{amssymb}
\usepackage{amsfonts} 
\usepackage{graphicx}
\usepackage{epsfig}
\usepackage{dcolumn}
\usepackage{multirow}
\usepackage{bm}
\usepackage{color}
\usepackage[ps2pdf,
bookmarks=true,%                   %%% generate bookmarks ...
bookmarksnumbered=true,%           %%% ... with numbers
hypertexnames=false,%              %%% needed for correct links to figures !!!
]{hyperref}
\usepackage{subfig}

\newcommand{\bra}[1]{\left< \smash{ #1 } \right|}
\newcommand{\ket}[1]{\left| \smash{ #1 } \right>}
\newcommand{\expect}[1]{\left< #1 \right>}

\newcommand{\mvec}[1]{\mathbf{ #1 }}

\newcommand{\abs}[1]{\left| #1 \right|}

\newcommand{\tr}[2][]{\text{Tr}_{\textsc{#1}} \left\{ \smash{ #2 } \right\}  }

\newcommand{\expo}[2][]{e^{#1 i \, #2}}
\newcommand{\polvec}[1]{{\bm{ \epsilon}}_{#1}}

\newcommand{\Aqu}[3][]{\hat{\mvec{A}}_{\textsc{#1}}^{#2}(#3)}
\newcommand{\Acl}[3][]{{\mvec{A}}_{\textsc{#1}}^{#2}(#3)}

\newcommand{\ZH}{Z}
\newcommand{\argline}[8]{(\smash{ \mvec{k}^{#1}_{#2}, {\omega}^{#3}_{#4}, \mvec{k}^{#5}_{#6}, {\omega}^{#7}_{#8}})}
\newcommand{\PProp}[1]{\mvec{P}_{#1}}
\newcommand{\negs}{\hspace{-1em}}
\newcommand{\intinf}{\int_{-\infty}^{\infty}}

\newcommand{\mytitle}{Theory of parametric x-ray optical wavemixing processes}

\newcommand{\rmpdfinfo}{\special{ps:: userdict /pdfmark /cleartomark load put}}

\definecolor{MyDarkGreen}{rgb}{0,0.6,0}
\definecolor{MyDarkBlue}{rgb}{0,0,0.8}
\definecolor{MyDarkRed}{rgb}{0.6,0,0.3}

\hypersetup{breaklinks=true,
colorlinks =true,
plainpages =true,
linktocpage=true,
linkcolor=MyDarkBlue,
citecolor=MyDarkGreen,
urlcolor =MyDarkRed,
pdfborder={0 0 0},
pdfauthor={Dietrich Krebs},
pdftitle ={\mytitle},
pdfsubject  = {Research Article},
pdfkeywords = {nonlinear Compton scattering},
pdfcreator  = {LaTeX with hyperref package},
pdfproducer = {dvipdf}
}

\begin{document} 
%%%%%%%%%%%%%%%%%%%%%%%%%%%%%%%%%%%%%%%%%%
%\setpagewiselinenumbers
%\modulolinenumbers[5]
%\linenumbers

\title{\mytitle}

% Author list 
\author{Dietrich Krebs}
\affiliation{Center for Free-Electron Laser Science, DESY, Notkestrasse 85, 22607 Hamburg, Germany}
\affiliation{Department of Physics, University of Hamburg, Jungiusstrasse 9, 20355 Hamburg, Germany}
\affiliation{The Hamburg Centre for Ultrafast Imaging, Luruper Chaussee 149, 22761 Hamburg, Germany}
\affiliation{Max Planck School of Photonics, Friedrich-Schiller University of Jena, Albert-Einstein-Str. 6, 07745 Jena, Germany}
%\affiliation{Malen nach Zahlen - Gesellschaft mit beschr\"ankter Hoffnung, c/o DESY, Notkestrasse 85, 22607 Hamburg, Germany}
%\thanks{}
\thanks{Corresponding author}
\email{dietrich.krebs@desy.de}

\author{Nina Rohringer}
\affiliation{Center for Free-Electron Laser Science, DESY, Notkestrasse 85, 22607 Hamburg, Germany}
\affiliation{Department of Physics, University of Hamburg, Jungiusstrasse 9, 20355 Hamburg, Germany}
\affiliation{The Hamburg Centre for Ultrafast Imaging, Luruper Chaussee 149, 22761 Hamburg, Germany}
\affiliation{Max Planck School of Photonics, Friedrich-Schiller University of Jena, Albert-Einstein-Str. 6, 07745 Jena, Germany}

\date{\today}

\begin{abstract}
Enabled by x-ray free-electron lasers, nonlinear optical phenomena can be explored in the x-ray domain nowadays.
Among the multitude of newly accessible processes, this theoretical study focuses parametric x-ray optical wavemixing for closer investigation. 
Specifically, we develop a framework based on non-relativistic QED to describe x-ray optical sum- and difference-frequency generation as well as x-ray parametric down-conversion on equal footing. 
All of these processes promise imaging capabilities similar to regular x-ray diffraction with additional spectroscopic selectivity that is tunable via the optical admixture.
Based on our derivation, we identify the imaged quantity as we relate the observable scattering pattern to an underlying response function of the medium. 
The resulting relation, furthermore, enables the microscopic reconstruction of this response function from nonlinear analogues of crystallographic measurements.
We benchmark our approach on recent experimental sum-frequency results, for which we find encouraging agreement with our theory.
\end{abstract}
  
%\pacs{ 32.80.Rm,42.65.Re,31.15.A-}
% 32.80.Rm  Multiphoton ionization and excitation to highly excited states 
% 42.65.Re  Ultrafast processes; 
%           optical pulse generation and pulse compression 
%           (for ultrafast spectroscopy, see 78.47.J-; 
%           for ultrafast magnetization dynamics, see 75.78.Jp)
% 31.15.A-  Ab initio calculations 
%\keywords{}
  
\maketitle
  
  
%%%%%%%%%%%%%%%%%%%%%%%%%%%%%%%%%%%%%%%%%%
\section{Introduction}
\label{sec:intro}
%%%%%%%%%%%%%%%%%%%%%%%%%%%%%%%%%%%%%%%%%%
Since their discovery in 1895, x-rays have been used to obtain insights into the structure of material systems---with x-ray diffraction in particular serving as a microscopic probe of the electronic density \cite{1912Laue-diffraction,1992Giacovazzo-BOOK}. This approach---backed by the continuous improvement of sources and detectors---has rendered it routinely possible, for example, to reconstruct complicated protein structures from x-ray scattering measurements nowadays \cite{2018Wiedorn_Chapman-MHz_SFX}.

Notwithstanding its broad success, regular x-ray diffraction becomes very challenging, if microscopic resolution of valence electron distributions is desired \cite{2005Coppens-Review}. These weakly bound electrons, which are the most relevant in chemical and technological terms, are typically spread dilutely across the solid and, as such, their scattering signal scales unfavorably compared to the well-localized (atomic) core electrons.

In order to circumvent this problem and obtain a valence selective scheme, Freund and Levine suggested to go beyond linear elastic scattering as early as 1970 \cite{1970Freund-optmodxrd}. Using parametric x-ray optical wavemixing processes (XOWM) instead, one can leverage the spectral selectivity of an additional optical field to isolate the response of valence electrons. 
An example of this is the process of sum-frequency generation (SFG), wherein a sample is illuminated by both x-ray ($\omega_X$) and optical ($\omega_O$) photons, while the detection focuses on scattered photons at the sum-frequency $\omega_S = \omega_X + \omega_O$. As this frequency cannot be generated via elastic scattering, any measured signal thus provides a short-wavelength 
probe \emph{conditional} on the simultaneous interaction with the optical modulation. Tuning the latter resonantly with valence excitations then establishes the desired selectivity.

Though promising in principle, x-ray wavemixing approaches were restricted to proof-of-principle tests for a long time, as they were stifled by low count rates and cumbersome detection \cite{1971Eisenberger-XPDCmeasured,1981Danino-XPDC335eV}. Yet, with the arrival of third and fourth generation light sources, they are now becoming realistically observable and have attracted renewed experimental interest \cite{2012Glover-SFG,2017Schori-opticalXPDC,2019Borodin-SharonPlasmon,2019Sofer-XPDCnaturecomm}. In fact, there has even been a first report addressing the long-standing goal to reconstruct the valence electron response \cite{2011Tamasaku-imagingchi1} within the high-frequency limit.

Complementing these experimental developments, we want to address the theoretical understanding of XOWM. So far, its interpretation often \cite{2012Glover-SFG,2017ZhengLi-GhostImagingXPDC,2017Borodin-PDCLabSource,2019Borodin-SharonPlasmon} relies on the classical picture suggested by Eisenberger and McCall for wavemixing within the x-ray domain \cite{1971Eisenberger-XPDCmeasured}. However, its central assumption---namely an electronic response in terms of free, classical currents---becomes invalid once optical frequencies are admixed. The optical response is instead dominated by discrete, quantum mechanical transitions that need to be accounted for in a theoretical description, in order to be microscopically accurate.
%properly allow for the reconstruction of microscopic material properties from XOWM signals.
Multiple attempts at this have been developed based on the semi-classical response formalisms of Armstrong et al. \cite{1962Armstrong-nonlinearoptics} as well as Jha et al. \cite{1968Jha-NonlinearResponseOneElectr}, with a noteworthy alternative recently presented by Gorelova et al., who treat the optical admixture via Floquet-formalism \cite{2018Gorelova-Floquet}. So far, none of these approaches has been evaluated beyond effective single electron theories (see also Sec.~\ref{sec:model}) \footnote{It should be noted that the stated theoretical shortcomings particularily concern XOWM in solid-state systems. For atomic and molecular systems, there is a host of well-developed nonlinear response formalisms that was advanced---among others---through the contributions of Refs.~\cite{1962Armstrong-nonlinearoptics,1968Jha-NonlinearResponseOneElectr}. Regarding x-ray optical SFG in molecules, for instance, a recent theoretical suggestion has been published by Rouxel et al. \cite{2018Rouxel-MukamelSFG}.}. 

In the present work, we derive a first principles description of XOWM processes---starting from nonrelativistic Quantum Electrodynamics (QED) \cite{1998CraigThirunamachandran-BOOK,1983Loudon-BOOK}. Our approach assumes an x-ray scattering perspective, the basis of which is outlined in Sec.~\ref{sec:theory}. Through it, we arrive at Eqs.~(\ref{eq:observable_fin_xt}) and (\ref{eq:observable_fin_kw}), which relate the scattering signal to separate correlation functions of the light-matter subsystems. Modelling these constituent functions in Sec.~\ref{sec:model} for the SFG scenario of Ref.~\cite{2012Glover-SFG}, we can ultimately benchmark our theory against said study in Sec.~\ref{sec:sfg}. We find encouraging agreement of theory and experiment, which suggests the viability of using the approach for structural reconstruction \footnote{\label{footnote-inverse-problem}We will discuss this inverse problem in more detail in a forthcoming publication \cite{future-meXXX}.}.

%old paragraph
% Based on these relations, measured XOWM signals could be inverted for structural reconstruction \footnote{We will discuss this inverse problem in more detail in a forthcoming publication \cite{future-me}.}. In the present discussion, however, we shall focus on the solution of the forward problem in order to benchmark the theory. We apply it to the SFG scenario of Ref.~\cite{2012Glover-SFG} (Sec.~\ref{sec:sfg}), for which we find good agreement between theoretical and experimental results. Modelling of the constituent correlation functions is detailed in the preceeding Secs.~\ref{ssec:matter} as well as \ref{ssec:light-x} and \ref{ssec:light-o} for the matter and light sub-systems, respectively.
 
Throughout the discussion, we will employ the conventional system of atomic units, i.e., \mbox{$\hbar = m_\text{e} = \abs{e} = 1$}, whereby we refer to the reduced Planck constant, the mass and the charge of an electron, respectively.
Furthermore, the speed of light in vacuo is referenced to $c = 1/\alpha$, with $\alpha$ denoting the fine structure constant.
When discussing structural properties of crystals (only cubic systems in this work), we refer to their conventional rather than their primitive lattices.

%collection 
%In our formulation of XOWM [Eqs.~(\ref{eq:observable_fin_xt}) or (\ref{eq:observable_fin_kw})], we relate the measurable scattering signal to separate correlation functions of the light and matter subsystems of the problem. This allows discussing the latter from the perspective of purely electronic response functions or electronic structure theory in general.

%%%%%%%%%%%%%%%%%%%%%%%%%%%%%%%%%%%%%%%%%%
\section{Theory}
\label{sec:theory}
%%%%%%%%%%%%%%%%%%%%%%%%%%%%%%%%%%%%%%%%%%

%%%%%%%%%%%%%%%%%%%%%%%%%%%%%%%%%%%%%%%%%%
\subsection{Basic framework}
\label{ssec:framework}
%%%%%%%%%%%%%%%%%%%%%%%%%%%%%%%%%%%%%%%%%%
We set our description of XOWM within the well-established framework of nonrelativistic QED \cite{1998CraigThirunamachandran-BOOK,1983Loudon-BOOK}, closely following the formalism outlined in Ref~\cite{2009Santra-Xraytutorial}. In particular, we choose to impose the Coulomb gauge upon the electromagnetic field and subsequently quantize only its transverse components. The respective vector potential in free space reads:
\begin{align}
  \Aqu{}{\mvec{x}}
  &=
  \sum_{\mvec{k},\lambda} \sqrt{\frac{2 \pi}{V \omega_\mvec{k} \alpha^2}} ~ \big( 
  \hat{a}_{\mvec{k},\lambda} ~\polvec{\mvec{k},\lambda} ~ \expo{\mvec{k}\cdot\mvec{x} } 
  + 
  \text{h.c.} 
  \big)
  ,
  \label{eq:em-pw-mode-decomp}
\end{align}
where $\hat{a}_{\mvec{k},\lambda}$ ($\hat{a}^{\dagger}_{\mvec{k},\lambda}$) is the annihilation (creation) operator for a photon of energy $\omega_\mvec{k} = c \, \abs{\mvec{k}}$ in the plane wave mode of wave vector $\mvec{k}$ and polarization $\polvec{\mvec{k},\lambda}$. There are two polarization states enumerated through $\lambda$, both of which satisfy the transversality condition $\polvec{\mvec{k},\lambda} \cdot \mvec{k} = 0$. The theoretical quantization volume is labelled by $V$. Notably, this volume will not feature in any physical observable (see below).
Coupling the field to matter, we focus on the electronic charges and write the associated interaction Hamiltonian in the familiar minimal coupling form \footnote{In writing $\hat{H}_\textsc{int}$ as of Eq.~(\ref{eq:H_int}), we neglect couplings of $\Aqu{}{\mvec{x}}$ to nuclear charges and---more intricately---to the spin degrees of freedom present in material systems.}:
\begin{align}
  \hat{H}_\textsc{int}
  &=
  \int \!\! d^3x \, \hat{\psi}^\dagger(\mvec{x}) \big( 
  \alpha \mvec{p} \cdot \Aqu{}{\mvec{x}}
  + \frac{\alpha^2}{2} \Aqu{2}{\mvec{x}}
  \big)  
  \hat{\psi}(\mvec{x})
  .
  \label{eq:H_int}
\end{align}
Here, the 2-component spinors $\hat{\psi}(\mvec{x})$ ($\hat{\psi}^\dagger(\mvec{x})$) are the electronic field operators analogous to $\hat{a}_{\mvec{k},\lambda}$ ($\hat{a}^{\dagger}_{\mvec{k},\lambda}$), which annihilate (create) an electron of particular spin state at the position $\mvec{x}$. We relegate any further details of the material's description (including electrostatic interactions) into a pertaining Hamiltonian $\hat{H}_\textsc{mat}$, while adopting the standard mode-based formulation for the electromagnetic part \mbox{$\hat{H}_\textsc{em} = \sum_{\mvec{k},\lambda} \omega_{\mvec{k}} (\hat{a}^{\dagger}_{\mvec{k},\lambda}\hat{a}_{\mvec{k},\lambda} + 1/2)$}.
Together with Eq.~(\ref{eq:H_int}), these render the Hamiltonian of the overall problem tripartite:
\begin{align}
  \hat{H} 
  &=
  \hat{H}_\textsc{mat}  + \hat{H}_\textsc{em} + \hat{H}_\textsc{int} 
  .
  \label{eq:H_overall}
\end{align}
Based on this, we can formally describe the time-evolution of the coupled system's density operator $\hat{\rho}(t)$ using the Liouville-von Neumann equation 
and, by extension, evaluate any physical observable as:
\begin{align}
  \langle \hat{O} \rangle (t)
  &=
  \tr{\hat{\rho}(t)\hat{O}}
  .  
  \label{eq:observable_reminder}
\end{align}
In particular, we shall proceed to define an operator $\hat{O}$ that suitably captures XOWM. To this end, we adopt a scattering perspective and focus on the detection of x-ray photons that are scattered out of an incoming field distribution into initially unoccupied modes of Eq.~(\ref{eq:em-pw-mode-decomp}). For any such final mode \smash{$\mvec{k}_f,\lambda_f$} we can determine the probability of a single photon being scattered into it by projecting onto the number state \smash{$\ket{1}_{\mvec{k}_f,\lambda_f}$}. Analogous procedures are widely used in the calculation of far-field diffraction patterns from weakly scattering samples (see for instance Refs.~\cite{2007Schuelke-BOOK,2011Als-Nielsen_BOOK-Modern_x-ray,2008Ho_Santra-spinning-tops,2015Slowik_PHD}).
More formally, we can write a projector across the full radiative Fock-space
\begin{align}
  \hat{\Pi}_{\mvec{k}_f, \lambda_f}
  &= 
% stacked indices below the product  
%  \big( \bigotimes_{\substack{\mvec{k} \neq \mvec{k}_f\\
%                              \lambda  \neq \lambda_f}  } \hat{\mathbf{1}}_{\mvec{k},\lambda} \big) \otimes \ket{1}\bra{1}_{\mvec{k}_f,%\lambda_f}
  \big( \bigotimes_{\mvec{k} \neq \mvec{k}_f, \lambda  \neq \lambda_f}   \negs \hat{\mathbf{1}}_{\mvec{k},\lambda} \big) \otimes \ket{1}\bra{1}_{\mvec{k}_f,\lambda_f}
  ,
  \label{eq:projection_observable_2}
\end{align}
which singles out the same probability.
Combining this with the identity on the material part, \mbox{\smash{$\hat{\Pi}_{\mvec{k}_f, \lambda_f} \otimes \hat{\mathbf{1}}_{\textsc{mat}} = \hat{O}_{\mvec{k}_f,\lambda_f}$}}, yields the desired observable for our purpose \footnote{It should be noted that---by construction---our observable is neither suited for cases of strong or (self-) stimulated scattering, where \smash{$\expect{\hat{n}_{\mvec{k}_f,\lambda_f}} \gtrsim 1$}, nor for cases of time-gated detection.}. 
In order to ultimately detach our results from the arbitrary quantization volume and its associated modes, we can convert the observable into a double differential scattering probability via the usual prescription \cite{2009Santra-Xraytutorial}:
\begin{align}
  \frac{\text{d} P_{\lambda_f}(\mvec{k}_f)}{d\Omega_f d\omega_f} 
  &= 
  \frac{V \alpha^3 }{(2 \pi)^3} \, \omega_f^2 \, \langle \hat{O}_{\mvec{k}_f,\lambda_f} \rangle
  .
  \label{eq:double_diff_prob}
\end{align}
\\
%%%%%%%%%%%%%%%%%%%%%%%%%%%%%%%%%%%%%%%%%%
\subsection{Perturbative approach to XOWM}
\label{ssec:pert}
%%%%%%%%%%%%%%%%%%%%%%%%%%%%%%%%%%%%%%%%%%
With the basic framework outlined, we return to Eq.~(\ref{eq:observable_reminder}) and concern ourselves in greater detail with the indicated time evolution of $\hat{\rho}(t)$. Its exact treatment is notoriously infeasible for realistic systems, such that we have to approach it via a series of simplifications.
Most notably, we choose to employ (conventional) time-dependent perturbation theory (see for example Ref.~\cite{1995PeskinSchroeder-BOOK})---assuming the light-matter coupling to be sufficiently weak---in order to decompose the fully interacting problem into its different subsystems.
To this end, we rewrite Eq.~(\ref{eq:observable_reminder})
\begin{align}
  \langle \hat{O}_{\mvec{k}_f,\lambda_f} \rangle (t)
  &=
  \tr{\hat{U}(t,t_0)\hat{\rho}(t_0)\hat{U}(t_0,t)\hat{O}_{\mvec{k}_f,\lambda_f}}
  \label{eq:observable_reminder_U}
\end{align}
utilizing the time-evolution operator $\hat{U}(t,t_0)$, which formally propagates the system from time $t_0$ to $t$.
We split this furter into $\hat{U}(t,t_0) = \hat{U}_\textsc{0}(t,t_0) \, \hat{U}_\textsc{int}(t,t_0)$, where each $\hat{U}_\textsc{...}$ is the propagator to a specific time dependent Schroedinger equation, involving $\hat{H}$, $\hat{H}_\textsc{0} = \hat{H} - \hat{H}_\textsc{int}$ or $\hat{H}_\textsc{int}$, respectively. 
Then---by expanding $\hat{U}_\textsc{int}(t,t_0)$ in powers of  $\hat{H}_\textsc{int}$---we arrive at the desired perturbation series.
We apply this procedure in two successive steps, focusing on the x-ray part of the interaction first and accounting for any optical admixture in a second iteration.

Preparing the first step, we formally partition the vector potential into orthogonal components
\begin{align}
  \Aqu{}{\mvec{x}}
  &= 
  \Aqu[x\_in]{}{\mvec{x}} + \Aqu[x\_out]{}{\mvec{x}} + \Aqu[opt]{}{\mvec{x}} + ...
  \label{eq:em-partitioning}
\end{align}
that pertain to incoming x-ray modes, scattered (or outgoing) x-ray modes as well as optical contributions to the em-field and a presently irrelevant remainder.
We employ this to rewrite Eq.~(\ref{eq:H_int}), keeping only terms that mediate x-ray scattering at lowest perturbative order of $\hat{U}_\textsc{int}(t,t_0)$. Therefore, we are left with
\begin{align}
  \hat{H}_\textsc{int\_x}
  &=
  \alpha^2 \int \!\! d^3x \, \hat{\psi}^\dagger(\mvec{x}) 
  \Aqu[x\_in]{}{\mvec{x}} \cdot \Aqu[x\_out]{}{\mvec{x}} \hat{\psi}(\mvec{x})
  ,
  \label{eq:H_int_x}
\end{align}
which corresponds to non-resonant ($A^2$) scattering and presents a suitable simplification if all involved photon energies ($\omega_\text{in}$, $\omega_\text{out}$) lie far off electronic resonances \cite{2015Slowik_PHD}.
We assume further that the initial conditions factorize
\begin{align}
  \hat{\rho}(t_0)
  &=
  \hat{\rho}_\textsc{x\_in}(t_0) \otimes \hat{\rho}_\textsc{x\_out}(t_0) \otimes \hat{\rho}_\textsc{sys}(t_0)
  ,
  \label{eq:factorize_x_rest}
\end{align}
i.e., that the incoming x-ray pulse, the initially unoccupied modes pertaining to scattered photons, and the (optically driven) material system are mutually uncorrelated prior to the scattering event.
This allows us to obtain the overall scattering observable in the compact form (for details see App.~\ref{sec:ixs}):
\begin{align}
  &\langle \hat{O}_{\mvec{k}_f,\lambda_f} \!\rangle 
  =
  \frac{2 \pi \alpha^2}{V \omega_f }  (\polvec{f})_\sigma (\polvec{f}^*)_\rho \!\!
  \int \!\! dt_1 \!\!\! \int \!\! dt_1^\prime \!\!\! \int \!\! d^3\!x \!\!\! \int \!\! d^3\!x^\prime  
  \expo{\omega_f (t_1-t_1^\prime)} \nonumber \\
  &\expo[-]{\mvec{k}_f (\mvec{x} - \mvec{x}^\prime)} \,
  (G_\textsc{x\_in}^{(1)}(\mvec{x}^\prime,t_1^\prime,\mvec{x},t_1))_{\sigma\rho} \,
  \langle \hat{n}(\mvec{x}^\prime,t_1^\prime) \, \hat{n}(\mvec{x},t_1) \rangle_\textsc{sys}
  %\tr[sys]{ \hat{n}(\mvec{x},t_1)  \hat{\rho}_\textsc{sys}(0) \hat{n}(\mvec{x}^\prime,t_1^\prime) }
  .
  \label{eq:observable_ixs_limit}
\end{align}
In writing Eq.~(\ref{eq:observable_ixs_limit}), we have made use of the following abbreviations: \smash{$(G_\textsc{x\_in}^{(1)})_{\sigma\rho}$} denotes the first-order correlation function of the incoming x-ray field, which is Glauber's usual coherence function \cite{1963Glauber-coherence} generalized to vector potentials:
\begin{align}
  &(G_\textsc{x\_in}^{(1)}(\mvec{x}^\prime,t_1^\prime,\mvec{x},t_1))_{\sigma\rho} \nonumber \\
  &=
  \tr[x\_in]{ 
  \hat{\rho}_\textsc{x\_in} \,
  (\Aqu[x\_in]{(-)}{\mvec{x}^\prime,t_1^\prime})_\sigma  
  (\Aqu[x\_in]{(+)}{\mvec{x},t_1})_\rho 
  }  
  .
\end{align}
%
 %; for our purposes, we define it directly on the level of vector potentials rather than the usual electric fields (see also maybe). %eg refer to the section where you discuss in more detail the modelling of light
We employ Einstein's summation convention on the polarization factors, where repeatedly occurring greek indices are understood to be summed over.
And, we include the correlator of two electron density operators $\hat{n}(\mvec{x})=\hat{\psi}^\dagger(\mvec{x}) \hat{\psi}(\mvec{x})$, each of which is written with a time-dependence according to \mbox{$\hat{n}(\mvec{x},t) = \hat{U}_\textsc{0}(0,t) \, \hat{n}(\mvec{x}) \, \hat{U}_\textsc{0}(t,0)$}. Their expectation value \mbox{$\langle ... \rangle_\textsc{sys} = \tr[sys]{ \hat{\rho}_\textsc{sys}(0) ... }$} is taken with respect to the system's initial state as of Eq.~(\ref{eq:factorize_x_rest})---correspondingly translated to $t=0$.

The obtained Eq.~(\ref{eq:observable_ixs_limit}) is consistent with established descriptions; its static limit rephrases usual inelastic x-ray scattering \cite{2007Schuelke-BOOK}, where the Fourier transformed density-density correlator is known as the dynamic structure factor. 
Conversely, in the case of time-resolved scattering, it reproduces---for instance---the results of Ref.~\cite{2012Dixit-TR-scattering}.
Considering more explicitly XOWM, it should further be noted that Eq.~(\ref{eq:observable_ixs_limit}) may already present the most suitable description for some scenarios---in particular if the optical light-matter coupling is strong and any further separation of $\hat{\rho}_\textsc{sys}(t)$ therefore unreasonable \footnote{Cases of strong optical light-matter coupling may include strong-field laser driving (cf., Ref.~\cite{2018Gorelova-Floquet}), but also embedding in a cavity with ensuing polaritonic effects.}.

Within the scope of this article, however, we want to assume that further separation is indeed possible. Hence, our second iteration of expansion aims at obtaining an expression for $\langle \hat{O}_{\mvec{k}_f,\lambda_f} \!\rangle$ that reflects the material's response to explicit x-ray and optical perturbations---highlighting the wave\emph{mixing} aspect.
In proceeding, we distinguish two facets of the optical light-matter interaction; The first concerns pure propagation effects inside the probed material (e.g., refraction), while the second is the actual non-linear wavemixing. We account for the former through our modelling of the incoming field (see Sec.~\ref{ssec:light-o}) \footnote{It should be noted that this dichotomy is regrettable with regard to the otherwise unified QED formalism. Presently, however, it may be justified by its practical viability, while a more formal justification in the spirit of leg-corrections \cite{1995PeskinSchroeder-BOOK} is relegated to future investigations.} and focus the perturbative expansion explicitly on the nonlinear process.
Starting from the density-density correlator in Eq.~(\ref{eq:observable_ixs_limit})
\begin{align}
  &\langle \hat{n}(\mvec{x}^\prime,t_1^\prime) \, \hat{n}(\mvec{x},t_1) \rangle_\textsc{sys} =  \label{eq:correlator} \\  
  &\tr[sys]{   
  \hat{\rho}_\textsc{sys}(0) \,
  \hat{U}_\textsc{0}(0,t_1^\prime) \,  \hat{n}(\mvec{x}^\prime) \, \hat{U}_\textsc{0}(t_1^\prime,t_1) \, \hat{n}(\mvec{x}) \, \hat{U}_\textsc{0}(t_1,0)
  }  \nonumber
\end{align}
the procedure is similar to the x-ray case.
We assume factorizing initial conditions in order to separate the optical and material subsystems of $\hat{\rho}_\textsc{sys}$. Subsequently, we rely on their weak interaction to expand all propagators $\hat{U}_\textsc{0}$ into further perturbative expressions.
In doing so, we restrict our analysis to lowest order contributions and---notably---only \emph{parametric} processes (for details see App.~\ref{sec:omix}).
The relevant interaction Hamiltonian under these conditions reads:
\begin{align}
  \hat{H}_\textsc{int\_opt}
  &=
  \alpha  \int \!\! d^3y \, \hat{\psi}^\dagger(\mvec{y})  \,  \mvec{p} \cdot \Aqu[opt]{}{\mvec{y}} \,  \hat{\psi}(\mvec{y})
  \label{eq:H_int_opt}
\end{align}
and the expansion of Eq.~(\ref{eq:correlator}) yields:
%%%
\begin{align}
  &\langle \hat{n}(\mvec{x}^\prime,t_1^\prime) \, \hat{n}(\mvec{x},t_1) \rangle_\textsc{sys} = \int \!\! dt_2 \!\!\! \int \!\! dt_2^\prime \!\!\! \int \!\! d^3y \!\!\! \int \!\! d^3y^\prime \,\nonumber \\
  &\alpha^2 \,   
  \tr[opt]{ 
  \hat{\rho}_\textsc{opt}(0) \, (\Aqu[opt]{}{\mvec{y}^\prime,t_2^\prime})_\nu \, (\Aqu[opt]{}{\mvec{y},t_2})_\mu 
  } \nonumber \\
  &\sum_{I} \, p_I 
  \left( \PProp{I} (\mvec{y},t_2, \, \mvec{x},t_1) \right)_\mu
  \,
  \left( \PProp{I} (\mvec{y}^\prime,t_2^\prime, \, \mvec{x}^\prime,t_1^\prime) \right)_\nu^*
  .
  \label{eq:tr_sys_resolve}
\end{align}
Here, we employ the abbreviation 
%%%
\begin{align}
  \left( \PProp{I} (\mvec{y},t_2, \, \mvec{x},t_1) \right)_\mu
  &=
  \bra{I} \, \hat{T}
  \left[(\hat{\mvec{p}}(\mvec{y},t_2))_\mu \, \hat{n}(\mvec{x},t_1)\right] \,
  \ket{I} 
  \label{eq:PProp}
\end{align}
to denote the time-ordered correlation function of the electronic system's density $\hat{n}(\mvec{x},t_1)$ with its momentum density $\hat{\mvec{p}}(\mvec{y},t_2) = \hat{\psi}^\dagger(\mvec{y},t_2)  \, \mvec{p} \,  \hat{\psi}(\mvec{y},t_2)$---measured in the state $\ket{I}$. 
The fact that either correlation function in Eq.~(\ref{eq:tr_sys_resolve}) also represents a transition amplitude from an initial state $\ket{I}$ into the same final state reflects the required parametric nature of XOWM. An additional weighting factor $p_I$ is included to account for the statistical mixture of multiple initial configurations. %Notably, the operators' time-dependence on the right hand side of Eq.~(\ref{eq:correlator_para}) differs from the previously employed and reads \mbox{$\hat{n}(\mvec{x},t) = \hat{U}_\textsc{00}(0,t) \, \hat{n}(\mvec{x}) \, \hat{U}_\textsc{00}(t,0)$}.

Having dealt with the matter subsystem, we turn our attention towards the correlator of optical fields in Eq.~(\ref{eq:tr_sys_resolve}). 
It would be appealing to replace this in terms of a Glauber-type correlation function again.
However, the expression at hand is more general than the previously encountered x-ray case. 
Here, we have not yet restricted our description to photon-emission or absorption (negative or positive frequency part of $\hat{\mvec{A}}_{\textsc{opt}}$, respectively).
Retaining this generality, we opt for a different reformulation: 
\begin{align}
  \Aqu[opt]{}{\mvec{y},t_2}
  &=
  \Acl[avg]{}{\mvec{y},t_2} + \Aqu[flu]{}{\mvec{y},t_2}
  .
  \label{eq:split_field}
\end{align}
We split the operator into a `classical' field $\Acl[avg]{}{\mvec{y},t_2} := \tr[opt]{\hat{\rho}_\textsc{opt}(0) \, \Aqu[opt]{}{\mvec{y},t_2} }$---defined via its average---and the remaining fluctuations.
This intrinsically exact separation lends itself straightforwardly to further approximations.
Whenever a spectral region is strongly affected by a classical, external field---for instance a driving laser---we resort to:
\begin{align}
  \tr[opt]{...}
  &\approx
  (\Acl[avg]{}{\mvec{y}^\prime,t_2^\prime})_\nu \, (\Acl[avg]{}{\mvec{y},t_2})_\mu \nonumber \\
  &=:
  (\bar{G}_\textsc{opt}^{(1)}(\mvec{y}^\prime,t_2^\prime,\mvec{y},t_2))_{\nu\mu}
  .
  \label{eq:tr_opt_drv}
\end{align}
Conversely, when such driving is absent, we expect any fluctuations to resemble the thermal equilibrium case and write:
\begin{align}
  \tr[opt]{...}
  &\approx
  \tr[opt]{
  \hat{\rho}_\textsc{opt}^{equ} (\Aqu[flu]{}{\mvec{y}^\prime,t_2^\prime})_\nu \, (\Aqu[flu]{}{\mvec{y},t_2})_\mu } \nonumber \\
  &=:
  (\bar{S}_\textsc{opt}^{(1)}(\mvec{y}^\prime,t_2^\prime,\mvec{y},t_2))_{\nu\mu}
  .
  \label{eq:tr_opt_equ}
\end{align}
Formally, we keep both options and combine Eqs.~(\ref{eq:tr_opt_drv}) and (\ref{eq:tr_opt_equ}) with our results from Eqs.~(\ref{eq:observable_ixs_limit}) and (\ref{eq:tr_sys_resolve}) into the overall XOWM observable
\begin{widetext}
\begin{align}
  &\langle \hat{O}_{\mvec{k}_f,\lambda_f} \!\rangle 
  =
  \frac{2 \pi \alpha^4}{V \omega_f }   \!\!
  \intinf \negs dt_1 \!\! \intinf \negs dt_1^\prime \!\! \intinf \negs dt_2 \!\! \intinf \negs dt_2^\prime \!\!\! 
  \int \!\! d^3\!x \!\!\! \int \!\! d^3\!x^\prime \!\!\! \int \!\! d^3y \!\!\! \int \!\! d^3y^\prime \,
  \expo{\omega_f (t_1-t_1^\prime)} \, \expo[-]{\mvec{k}_f (\mvec{x} - \mvec{x}^\prime)} \, (\polvec{f})_\sigma (\polvec{f}^*)_\rho \,
  (G_\textsc{x\_in}^{(1)}(\mvec{x}^\prime,t_1^\prime,\mvec{x},t_1))_{\sigma\rho} \nonumber \\
  &\Big(
  (\bar{G}_\textsc{opt}^{(1)}(\mvec{y}^\prime,t_2^\prime,\mvec{y},t_2))_{\nu\mu} + (\bar{S}_\textsc{opt}^{(1)}(\mvec{y}^\prime,t_2^\prime,\mvec{y},t_2))_{\nu\mu}
  \Big) \,  
  \sum_{I} \, p_I 
  \left( \PProp{I} (\mvec{y},t_2, \, \mvec{x},t_1) \right)_\mu
  \,
  \left( \PProp{I} (\mvec{y}^\prime,t_2^\prime, \, \mvec{x}^\prime,t_1^\prime) \right)_\nu^*
  .
  \label{eq:observable_fin_xt}
\end{align}
%2column
%\begin{align}
%  &\langle \hat{O}_{\mvec{k}_f,\lambda_f} \!\rangle 
%  =
%  \frac{2 \pi \alpha^4}{V \omega_f }   \!\!
%  \int \!\! dt_1 \!\!\! \int \!\! dt_1^\prime \!\!\! \int \!\! dt_2 \!\!\! \int \!\! dt_2^\prime \!\!\! 
%  \int \!\! d^3\!x \!\!\! \int \!\! d^3\!x^\prime \!\!\! \int \!\! d^3y \!\!\! \int \!\! d^3y^\prime \,\nonumber \\
%  %
%  &\expo{\omega_f (t_1-t_1^\prime)} \, \expo[-]{\mvec{k}_f (\mvec{x} - \mvec{x}^\prime)} \, (\polvec{f})_\sigma (\polvec{f}^*)_\rho \,
%  (G_\textsc{x\_in}^{(1)}(\mvec{x}^\prime,t_1^\prime,\mvec{x},t_1))_{\sigma\rho} \nonumber \\
%  %
%  &\Big(
%  (\bar{H}_\textsc{opt}^{(1)}(\mvec{y}^\prime,t_2^\prime,\mvec{y},t_2))_{\nu\mu} + (\bar{S}_\textsc{opt}^{(1)}(\mvec{y}^\prime,t_2^\prime,\mvec{y},t_2))_{\nu\mu}
%  \Big)  \nonumber \\
%  &\sum_{I} \, p_I 
%  \left( \PProp{I} (\mvec{y},t_2, \, \mvec{x},t_1) \right)_\mu
%  \,
%  \left( \PProp{I} (\mvec{y}^\prime,t_2^\prime, \, \mvec{x}^\prime,t_1^\prime) \right)_\nu^*
%  .
%  \label{eq:observable_fin_xt}
%\end{align}
%\end{widetext}
%

While the result above is formulated in real space and time, it will prove advantageous---in particular for the analysis of crystalline matter---to transform into reciprocal coordinates.
To this end, we substitute all occurring quantities by their Fourier transforms (for details see App.~\ref{sec:Fourier}) and resolve the arising $\delta$-function constraints. Thereby, we obtain
%
%\begin{widetext}
\begin{align}
  &\langle \hat{O}_{\mvec{k}_f,\lambda_f} \!\rangle
  = 
  \frac{\alpha^4}{V \omega_f \,(2\pi)^{13}} \,
  \intinf \negs d\omega_O \!\! \intinf \negs d\omega_O^\prime \!\! \intinf \negs d\omega_X \!\! \intinf \negs d\omega_X^\prime \!\!\!
   \int \!\! d^3k_O \!\!\! \int \!\! d^3k_O^\prime \!\!\! \int \!\! d^3k_X \!\!\! \int \!\! d^3k_X^\prime \,  
  \delta(\omega_f - \omega_X - \omega_O) \, \delta(\omega_f - \omega_X^\prime - \omega_O^\prime)  \nonumber \\
  &(\polvec{f})_\sigma \, ({\ZH}_{\textsc{x\_in}} \argline{\prime}{X}{\prime}{X}{}{X}{}{X})_{\sigma\rho} \, (\polvec{f}^*)_\rho \,
  \Big( (\bar{\ZH}_{\textsc{opt}} \argline{\prime}{O}{\prime}{O}{}{O}{}{O})_{\nu\mu} + (\bar{C}_{\textsc{opt}} \argline{\prime}{O}{\prime}{O}{}{O}{}{O})_{\nu\mu} \Big) \nonumber \\
  &\sum_I \, p_I \, 
  \left( \mvec{K}_I(-\mvec{k}_O,\mvec{k}_f-\mvec{k}_X,\omega_O) \right)_\mu
  \left( \mvec{K}_I(-\mvec{k}_O^\prime,\mvec{k}_f-\mvec{k}_X^\prime,\omega_O^\prime) \right)^*_\nu  
  .
  \label{eq:observable_fin_kw}
\end{align}
\end{widetext}
Notably, we have kept the expression $\delta(\omega_f - \omega_X - \omega_O)$ from being integrated as it highlights two important aspects:

First, we can observe that the overall expression indeed describes \emph{parametric} XOWM. Frequency components of the optical field ($\omega_O$) are mixed with x-ray contributions ($\omega_X$) to produce the outgoing photon's frequency $\omega_f$---involving \emph{no} net exchange of energy with the material system.

Second, depending on the choice of $\omega_f$ relative to the central frequency components of incoming x-ray and optical fields, Eq.~(\ref{eq:observable_fin_kw}) describes either sum- or difference frequency generation. If no optical field is applied externally, the expression can likewise be employed to describe x-ray parametric down-conversion (XPDC). Thus, Eq.~(\ref{eq:observable_fin_kw}) provides access to a broad range of parametric XOWM processes that are relevant to date on a coherent footing.

%%%%%%%%%%%%%%%%%%%%%%%%%%%%%%%%%%%%%%%%%%
\section{Modeling}
\label{sec:model}
%%%%%%%%%%%%%%%%%%%%%%%%%%%%%%%%%%%%%%%%%%
%
In order to evaluate Eq.~(\ref{eq:observable_fin_kw}) quantitatively, we have to understand its constituent correlation functions and formulate appropriate models for them.
Aiming to benchmark our overall approach on the SFG results by Glover et al. \cite{2012Glover-SFG}, we naturally choose our description of the incident x-ray and optical fields to reflect their experimental conditions (see Secs.~\ref{ssec:light-x} and \ref{ssec:light-o}, respectively).
In addition, we require knowledge of the nonlinear response function $\mvec{K}_I$ for the case of a Diamond sample.
The latter is a material specific property and will subsequently be discussed in greater detail.

%%%%%%%%%%%%%%%%%%%%%%%%%%%%%%%%%%%%%%%%%%
\subsection{Electronic response}
\label{ssec:matter}
%%%%%%%%%%%%%%%%%%%%%%%%%%%%%%%%%%%%%%%%%%
%
%
The essential material quantity in our formulation of XOWM is the electronic response function $\PProp{I}(\mvec{y},t_2, \, \mvec{x},t_1)$---or its Fourier transform $\mvec{K}_I(\mvec{k}_1,\mvec{k}_2,\omega)$, respectively.
Having separated this from the radiative aspects of the process, we can proceed to review it from a perspective of general electronic structure or response theory.
As such, we established in Eq.~(\ref{eq:PProp}) that $\PProp{I}$ is the time-ordered correlator of the material's electronic density $\hat{n}(\mvec{x},t_1)$ and electronic momentum density $\hat{\mvec{p}}(\mvec{y},t_2)$. 
Structurally, this correlator is a causal, two-particle Green's function---also referred to as polarization propagator \cite{1984Oddershede-PolProp}.
It arises naturally in the context of light-matter interaction, where it forms part of a larger ($4 \times 4$) response tensor interlinking density and current-density response functions \cite{2013stefanucci_vanleeuwen-BOOK}.
Considering this response for electronic many-body systems, $\PProp{I}$ will naturally reflect their correlation properties.
By preserving this important aspect, our formulation of XOWM enables accurate forward calculations of scattering pattern based on established techniques from many-body theory (cf., for instance, Refs.~\cite{1962Kadanoff_Baym_BOOK,1990Gross-TDDFT_1st,2002Rubio-TDDFT_and_resp_fct,2006Kotliar_DMFT_review,2013Booth_FCIQMC,2020Wang_exciton_CCSD}).
Moreover, it provides the basis to extract such properties from XOWM measurements on correlated systems---upon solving the inverse problem \footnotemark[1].

In contrast, previous approaches to describe XOWM, remained restricted to \emph{single-particle} treatments. The perturbative formulations of this problem fell roughly into two classes.
The first comprises discussions such as Refs.~\cite{2012Glover-SFG,2017ZhengLi-GhostImagingXPDC,2017Borodin-PDCLabSource,2019Borodin-SharonPlasmon}, adopting the classical current-based picture that was introduced by Eisenberger and McCall for wavemixing within the x-ray domain \cite{1971Eisenberger-XPDCmeasured}.
This captures the nonlinear electronic response at high frequencies (cf. Ref.~\cite{1999Adams-XPDC}), where individual electrons interact with the driving fields like free, classical charges. 
However, this picture is not immediately transferable towards lower photon energies---especially if optical admixtures are involved. Then, the bound nature of the electrons becomes essential as their response is dominated by discrete, quantum-mechanical transitions.
This aspect is taken more prominently into account by the second class of previously published XOWM descriptions \cite{1970Freund-optmodxrd,1972Freund-nonlindiffr,1971Eisenberger-Mixing,1972Woo-IXS_optically_induced,1972Jha-nonlinearXray,1972_Jha_NuovoCim-NonlinearXrays,2019Cohen-Sharontheory}. Herein, the quantum-mechanical nature of the response is related to single-particle wavefunctions featuring in perturbative expressions.
All of these approaches can be traced to the semi-classical response formalisms presented by Armstrong et al. \cite{1962Armstrong-nonlinearoptics} as well as Jha et al. \cite{1968Jha-NonlinearResponseOneElectr}.
Despite being half a century old though, these expressions have never been evaluated in their full form for XOWM. At most, Freund has presented results for the simplified case of a ``high-frequency'' limit \cite{1972Freund-nonlindiffr}, while in addition, bond-charge models \cite{1970Freund-optmodxrd} and toy-model semiconductors \cite{2019Cohen-Sharontheory} have been considered \footnote{It should be noted that---outside the scope of models presented here---there was a single publication by Van Vechten and Martin that also alludes to XOWM in the context of calculating local field effects in solids \cite{1972vanVechten-LocalFieldPDC}. They conduct a full numerical evaluation of the Adler-Wiser dielectric response in random-phase approximation \cite{1962Adler-Dielectric,1963Wiser-dielectric} and present a result to compare with Ref.~\cite{1970Freund-optmodxrd}. Unfortunately, they do not give details on their reasoning, though their result likely presents the most elaborate computation on perturbative XOWM to-date.}.

For our present benchmark, instead, we pursue an evaluation of the material's response based on first principles. To this end, we employ a mean-field description of the electronic structure---more specifically Density-Functional-Theory (DFT) \cite{1964Hohenberg_Kohn_DFT,1965Kohn_Sham_DFT}---and express the field operators inherent in expression (\ref{eq:PProp}) in terms of Kohn-Sham orbitals \footnote{It should be noted that using a mean-field description amounts to adopting an \emph{effective} single particle picture. Nonetheless, this follows consistently from the many-body expression and thus provides a means to approximate the generally complex problem.}:
\begin{align}
%  \PProp{I} (\mvec{y},t_2, \, \mvec{x},t_1) 
%  &=
%  2 \, \sum_i^{\text{occ.}} \, \sum_a^{\text{unocc.}} \, \Big( 
%  \Theta(t_1-t_2) ~ \expo{(\varepsilon_i - \varepsilon_a)(t_1-t_2)} ~ \mvec{M}_{i,a}(\mvec{x},\mvec{y})
%  +
%  \Theta(t_2-t_1) ~ \expo[-]{(\varepsilon_i - \varepsilon_a)(t_1-t_2)} ~ (\mvec{M}_{i,a}(\mvec{x},\mvec{y}))^*
%  \Big)
%  .
  \PProp{I} (\mvec{y},0, \, \mvec{x},t) 
  &=
  2 \, \sum_i^{\text{occ.}} \sum_a^{\text{unocc.}} \, \Big( 
  \Theta(t) \, \expo{(\varepsilon_i - \varepsilon_a)t} ~ \mvec{M}_{i,a}(\mvec{x},\mvec{y}) \nonumber \\
  &+
  \Theta(-t) \, \expo[-]{(\varepsilon_i - \varepsilon_a)t} ~ \mvec{M}_{a,i}(\mvec{x},\mvec{y})
  \Big)
  .
\end{align}
Here, we have used the fact that $\PProp{I}$ is invariant under time-translations and condensed its prior arguments \mbox{$t_1 - t_2 = t$}. The matrix elements $\mvec{M}_{i,a}$ are given by
\begin{align}
  \mvec{M}_{i,a}(\mvec{x},\mvec{y})
  &=
  \varphi_i^*(\mvec{x}) \, \varphi_a(\mvec{x}) \, \varphi_a^*(\mvec{y}) \, (-\text{i} \nabla) \, \varphi_i(\mvec{y})
\end{align}
with $\varphi_p(\mvec{x})$ denoting said orbitals and $\varepsilon_p$ the corresponding eigenenergies with respect to the mean-field Hamiltonian.
We employ indices $i,j$ when referring to occupied orbitals (``valence bands'') and indices $a,b$ in case of unoccupied orbitals (``conduction band'')---noting that this clear separation is particular to simple, gapped materials, which admit a single-reference description. 

In view of the numerical evaluation, %\cite{2020Gonze-Abinit}
we adapt $\PProp{I}$ from an infinite, continuous domain onto a finite volume $V_{\diamond}$ with periodic boundary conditions \cite{1912Born-BvK}:
\begin{align}
  \PProp{I} (\mvec{y},0, \, \mvec{x},t)
  &\approx
  w(\mvec{y}) ~ w(\mvec{x}) ~  
  \PProp{I\diamond} (\mvec{y},0, \, \mvec{x},t) 
  .
  \label{eq:form-funtion-corr}
  \end{align}
We account for the proper extent of the sample by means of a form function that is $w(\mvec{x}) = 1$ inside (and $w(\mvec{x}) = 0$ outside) the material, whereby we follow a common practice in regular crystallography \cite{1992Giacovazzo-BOOK}.
Taking the Fourier transform of the new $\PProp{I\diamond}$, we obtain
\begin{align}
  \mvec{K}_{I\diamond}(&\mvec{q} + \mvec{G}_1, -\mvec{q} + \mvec{G}_2,\omega)
  =
  \lim\limits_{\epsilon \to 0^+} ~
  2 \text{i} \, \sum_i^{\text{occ.}} \, \sum_a^{\text{unocc.}} \,  \label{eq:e-corrf-mf-full} \\
  \Big(
  &\frac{\bra{\varphi_i} \expo[-]{(-\mvec{q} + \mvec{G}_2)\cdot\hat{\mvec{x}}} \ket{\varphi_a} \, \bra{\varphi_a} \expo[-]{(\mvec{q} + \mvec{G}_1)\cdot\hat{\mvec{x}}} \, \hat{\mvec{p}} \ket{\varphi_i} }{\omega - (\varepsilon_a - \varepsilon_i) + \text{i} \epsilon} \nonumber \\
  +
  &\frac{\bra{\varphi_i} \expo[-]{(\mvec{q} + \mvec{G}_1)\cdot\hat{\mvec{x}}} \, \hat{\mvec{p}} \ket{\varphi_a} \, \bra{\varphi_a} \expo[-]{(-\mvec{q} + \mvec{G}_2)\cdot\hat{\mvec{x}}} \ket{\varphi_i}}{-\omega - (\varepsilon_a - \varepsilon_i) + \text{i} \epsilon}
  \Big) 
  , \nonumber
\end{align}
with the involved time-integrals regularized through the prescription of adiabatic switching \cite{2009Santra-Xraytutorial}.
The spatial integrations on $V_{\diamond}$ are abbreviated into Dirac bra(c)kets of the involved orbitals.
Overall, Eq.~(\ref{eq:e-corrf-mf-full}) assumes the familiar structure of second-order perturbative expressions---featuring poles at transition energies and matrix elements weighing these transitions.
In particular, $\mvec{K}_{I\diamond}(\mvec{q} + \mvec{G}_1, -\mvec{q} + \mvec{G}_2,\omega)$ bears resemblance to the well-known Adler-Wiser response function \cite{1962Adler-Dielectric,1963Wiser-dielectric}, which is ubiquitously used in density-response formalisms---albeit without the transverse nature here implied by $\hat{\mvec{p}}$.

In order to simplify Eq.~(\ref{eq:e-corrf-mf-full}) further, we consider the typical values of $\mvec{q}$ and $\mvec{G}_1$. 
We can infer from Eq.~(\ref{eq:observable_fin_kw}) that the first argument of $\mvec{K}_{I}(\mvec{q} + \mvec{G}_1, -\mvec{q} + \mvec{G}_2,\omega)$ corresponds to the wave-vector of the admixed optical field. As such, it is much smaller than any reciprocal lattice vector, i.e., $\mvec{G}_1 = 0$. 
The remaining momentum transfer $\mvec{q} \sim \mvec{k}_O$ may moreover be neglected while evaluating the transition matrix elements.
This corresponds to treating optical interband transitions as effectively vertical in the Brillouin-zone.
Carrying these constraints over to the form function treatment of Eq.~(\ref{eq:form-funtion-corr}), we find its simplified correspondence in reciprocal space:
\begin{align}
  \mvec{K}_{I}(\mvec{k}_1, \mvec{k}_2,\omega) 
  &\approx
  \sum^{{\text{rec. latt.}}}_{\mvec{G}} \, \tilde{w}(\mvec{k}_1 + \mvec{k}_2 - \mvec{G}) \,
  \frac{1}{V_\diamond} \, \mvec{K}_{I\diamond}(\mvec{0}, \mvec{G},\omega) 
  \label{eq:transfer-K-to-pbc-dipole-app-fin}
  .
  \end{align}
Herein, $\tilde{w}(\mvec{k})$ marks the Fourier transform of ${w}(\mvec{x})$ and typically selects a single dominant contribution from the sum.

This being established, we turn to a numerical framework---in our case \textsc{abinit} \cite{2020Gonze-Abinit}---to supply the electronic structure input for the simplified correlation function.
In order to account for the notoriously wrong band-gap energy obtained from LDA-DFT \cite{1985Perdew-bandgaperror}, we employ a scissor correction of $\Delta E = 0.062 \text{ a.u. } (\sim 1.7 \text{ eV})$ adopted from Ref.~\cite{2004Botti-scissor}---for further parameters see \footnote{For the LDA-DFT calculations that provide the Kohn-Sham orbitals, we employ a norm-conserving pseudopotential and a plane-wave basis; the cut-off energy is taken to be $15 \text{ a.u.}$. The real space lattice of diamond is fixed to the cubic unit cell size of $6.741 \text{ a.u.}$, while the Brillouin zone is initially sampled at 60 k-points (symmetry adapted) to iterate the DFT calculation. Ultimately, the result is extrapolated onto 2048 k-points spread homogeneously across the full Brillouin zone (without symmetry adaption). The orbital energies of all unoccupied states are shifted by a scissor correction of $\Delta E = 0.062 \text{ a.u. } (\sim 1.7 \text{ eV})$ adopted from Ref.~\cite{2004Botti-scissor}.}.
Using a regularization $\epsilon = 0.007 \text{ a.u. } (\sim 0.2 \text{ eV})$ that paralles earlier, all-optical calculations on diamond \cite{1998Benedict-dielectric}, we proceed to compute  $\mvec{K}_{I\diamond}$. 
For the (111) reflection that was studied by Glover et al., we find our vectorial correlation function to be fully aligned with $\mvec{G}_{111}$.
Notably, this orientation already entails the correct polarization dependence of SFG (cf. Sec.~\ref{sec:sfg} and Fig.~\ref{fig:Glover-fake} (c)).

\begin{figure}[h!]
\begin{center}
  \rmpdfinfo
  \includegraphics[width=0.9\linewidth]{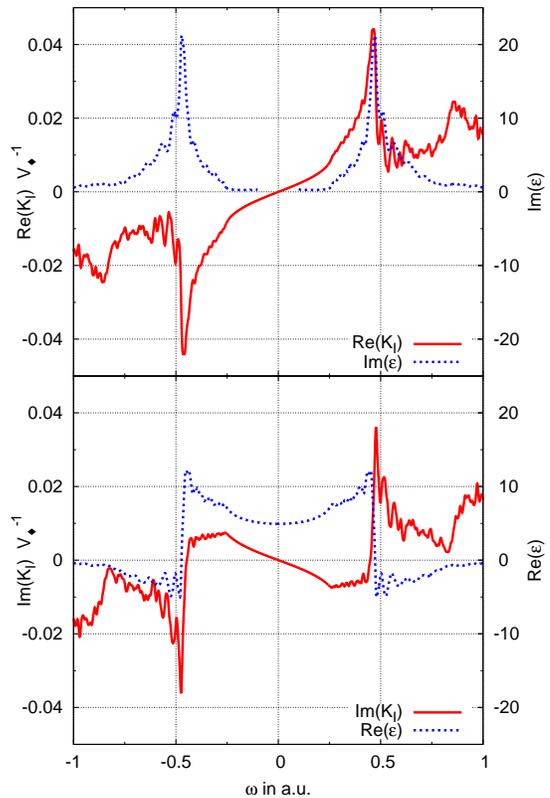}
  \caption{(Color online) The real and imaginary part of the nonlinear response are given as red solid lines in the upper and lower section of the graph, respectively. Specifically, $\mvec{K}_{I\diamond}(0, \mvec{G}_{111},\omega)$ is shown as obtained from our DFT-approach---projected onto $\mvec{G}_{111}/\abs{\mvec{G}_{111}}$ and normalized by $V_\diamond$. The imaginary and real part of the dielectric function, computed via the Kubo-Greenwood method \cite{2019Ramakrishna-KuboGreenwood} from the same DFT-structure, are juxtaposed as blue dashed lines, illustrating their spectral similarities. 
  }
  \label{fig:comp-K-vs-eps}
\end{center}
\end{figure}

Considering the spectral dependence of $\mvec{K}_{I\diamond}(0,\mvec{G}_{111},\omega)$, we plot the real and imaginary part of its projection onto $\mvec{G}_{111}/\abs{\mvec{G}_{111}}$ in Fig.~\ref{fig:comp-K-vs-eps} (upper and lower part, respectively). The data is normalized with respect to the volume of the simulated crystal domain $V_\diamond$ in line with Eq.~(\ref{eq:transfer-K-to-pbc-dipole-app-fin}).
For illustration and comparison, we also show the dielectric function as obtained via the Kubo-Greenwood formula \cite{2019Ramakrishna-KuboGreenwood} from the same Kohn-Sham orbitals. Its data is mirrored for negative frequencies.
Given their common excitation poles, we unsurprisingly find similar spectral features in both---prominently around $0.5 \text{ a.u.}$---with the real part of $\mvec{K}_{I\diamond}$ resembling the imaginary part of $\varepsilon(\omega)$ and vice-versa.
Towards higher energies, the spectral structure of the correlation function appears more pronounced than its dielectric counterpart.

Regarding the accuracy of either function, the limitations of ground state DFT should be borne in mind \cite{2016FerreBOOK-ExcitedStateDFT}.
Just as results for the dielectric function improve with the use of higher-level methods \cite{2019Ramakrishna-KuboGreenwood}, $\mvec{K}_{I\diamond}$ is likewise expected to evolve---incorporating excitonic and collective effects \footnote{It should be noted with regard to Plasmons as a possible collective effect that these concern a different part of the electromagnetic interaction than the correlation function $\mvec{K}_{I\diamond}$. While Plasmons are a longitudinally-coupling phenomenon, the correlation function pertains to transverse coupling. It is therefore questionable, whether plasmonic effects could be visible in XOWM at all. Recent experimental claims to this effect \cite{2019Borodin-SharonPlasmon}, thus call for careful examination.}.

\begin{figure}[h!]
\begin{center}
  \rmpdfinfo
  \includegraphics[width=0.9\linewidth]{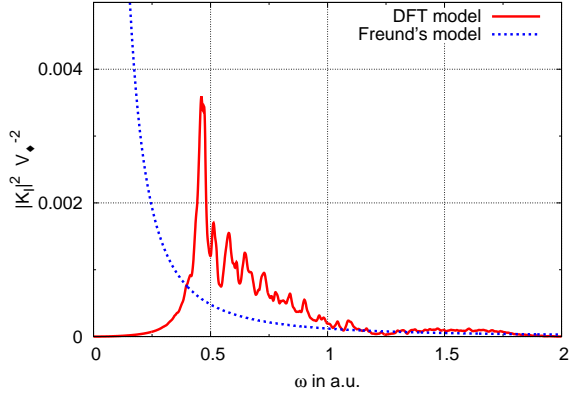}
  \caption{(Color online) $\abs{\mvec{K}_{I\diamond}(0, \mvec{G}_{220},\omega)}^2 V^{-2}_\diamond$ is computed based on two different approaches. The red solid line gives the DFT-based version stemming from Eq.~(\ref{eq:e-corrf-mf-full}), while the blue dashed line pertains to the approximate approach by Freund \cite{1972Freund-nonlindiffr} implemented via Eq.~(\ref{eq:e-corrf-Freund}).
  }
  \label{fig:comp-K-vs-KFreund}
\end{center}
\end{figure}

For an additional cross-check of our first-principles correlation function, we can compare to Freund's approach in the ``high-frequency'' limit \cite{1972Freund-nonlindiffr}.
Translating his assumptions and approximations into our formulation of  $\mvec{K}_{I\diamond}$, we can reduce it to the expression:
\begin{align}
  \mvec{K}_{I\diamond}(0, \mvec{G},\omega)
  &\approx
  \frac{2 \text{i}}{\omega} ~ \mvec{G} \, \int_\diamond d^3x ~ \rho_{\textsc{val}}(\mvec{x}) \, \expo[-]{\mvec{G}\cdot{\mvec{x}}} 
  \label{eq:e-corrf-Freund}
 .
\end{align}
This should give a reasonable estimate, if $\omega$ is much larger than typical valence-electron energy scales---yet much smaller than respective core-electron thresholds. Under these conditions, XOWM processes become directly sensitive to the valence-electron density $\rho_{\textsc{val}}(\mvec{x})$ of the sample. While such a clear probe would be desirable, the necessary conditions may \emph{not} be achievable for many realistic systems \cite{1972Freund-nonlindiffr}. In addition, it should be noted that the rate of asymptotic convergence depends on the reciprocal component in question.
Nevertheless, it is reassuring to find our full correlation function to be in good asymptotic agreement with the approximate Eq.~(\ref{eq:e-corrf-Freund})---for instance for the (220) direction in diamond, for which we compare spectra in Fig.~\ref{fig:comp-K-vs-KFreund}.
Hereunto, we plot the absolute squares of either function, as these give a measure of the scattering yields of interest.
We observe that Freund's approach provides a good fit above $1 ~\text{a.u.}~ (\approx 27.2\text{ eV})$ and overall a decent order of magnitude estimate. Nevertheless, it is obviously inadequate for determining any spectral features.
This highlights the necessity to develop more complete theoretical treatments, in order to describe XOWM accurately.

%%%%%%%%%%%%%%%%%%%%%%%%%%%%%%%%%%%%%%%%%%
\subsection{X-ray field}
\label{ssec:light-x}
%%%%%%%%%%%%%%%%%%%%%%%%%%%%%%%%%%%%%%%%%%
%
%could also cite singer's article in schneider's book : 2016_Schneider-BOOK_synchrotronAndFELs
%
The dependence of XOWM on the incident x-ray field is captured by the first order correlation function $(G_\textsc{x\_in}^{(1)})_{\sigma\rho}$ or its Fourier transformed counterpart $({\ZH}^{(1)}_{\textsc{x\_in}})_{\sigma\rho}$, respectively.
Its concrete form results from properties of the source, the manipulation of x-rays along the beam path and their propagation through the sample itself.
For synchrotron or free-electron laser (FEL) sources, certain elements of $(G_\textsc{x\_in}^{(1)})_{\sigma\rho}$ have been measured \cite{2012Singer-thesis,2011Vartanyants-CohLCLS} while others become accessible through simulations \cite{2013Chubar-SRWWavefront,2014Klementiev-XRT,2016Samoylova-WPG} or via modeling \cite{2010Geloni-FELStatistic,2015Kim-Synchrotron}.

For our current demonstration of XOWM, we choose a model-based approach that remains analytically treatable---avoiding the numerical expenses of accounting for the formally 8-dimensional dependence of $(G_\textsc{x\_in}^{(1)})_{\sigma\rho}$.
Specifically, we adopt a Gaussian Schell-model (GSM) \cite{1961Schell-AntennaThesis}, which was found to yield a good description of undulator based x-ray sources \cite{2012Singer-thesis,2017Gorobtsov-SingerCoh}.
Starting from Eqs.~(13) and (14) of Ref.~\cite{2017Gorobtsov-SingerCoh}, we similarily write the cross spectral density of a---notably still scalar---field as:
%%%
\begin{align}
  %W_\textsc{x\_in}^{(1)}((\mvec{x}^{\bot}_1,z_0),\omega_1,(\mvec{x}^{\bot}_2,z_0),\omega_2)
  W_\textsc{x\_in}^{(1)}(\mvec{x}_1,\omega_1,\mvec{x}_2,\omega_2) \Big|_{\substack{{z_1=z_0} \\ {z_2=z_0}}}
  &\propto
  J(\mvec{x}^{\bot}_1,\mvec{x}^{\bot}_2)  W(\omega_1,\omega_2)
  .
  \label{eq:crossspec-13}
\end{align}
The expression refers to a virtual source plane of the radiation at $z_0$, wherein the spatial and spectral components respectively read:
%%%
\begin{align}
  J(\mvec{x}^{\bot}_1,\mvec{x}^{\bot}_2) 
  =
  \text{exp}\Big[ &- \frac{(\mvec{x}^{\bot}_1)^2 + (\mvec{x}^{\bot}_2)^2}{4 \sigma^2} - \frac{(\mvec{x}^{\bot}_1 - \mvec{x}^{\bot}_2)^2}{2\xi^2} \Big] \label{eq:crossspec-14-0}\\
  W(\omega_1,\omega_2)
  =
  \text{exp}\Big[ &- \frac{(\omega_1 - \omega_0)^2 + (\omega_2 - \omega_0)^2}{4\Omega^2} \nonumber \\
                  &- \frac{(\omega_1 - \omega_2)^2}{2\Omega_c^2} - i t_0 (\omega_1 - \omega_2)\Big]
  .
  \label{eq:crossspec-14}
\end{align}
The coordinates $\mvec{x}^{\bot}_i$ span this plane perpendicular to the $z$-direction, which we will---in turn---adopt as the direction of propagation for our x-ray beam. For simplicity, we assume the initial distribution of radiation to be radially symmetric with a transverse size $\sigma$ and transverse coherence length $\xi$.  
Around its central frequency $\omega_0$, our model x-ray pulse features a bandwidth $\Omega = \sqrt{\smash{\tau_c}^{-2} + (2T)^{-2}}$ that is largely determined by the longitudinal coherence time $\tau_c$ and typically to a lesser extent by the average pulse length $T$.
Conversely, the spectral coherence width $\Omega_c = T^{-1} \sqrt{1 + \tau_c^2 (2T)^{-2}}$ mostly reflects the overall pulse length.
Adding to the description from Ref.~\cite{2017Gorobtsov-SingerCoh}, we include a phase-factor that accounts for the temporal center of the pulse being set to $t_0$.

In order to relate the above to $({\ZH}^{(1)}_{\textsc{x\_in}})_{\sigma\rho}$---as needed of Eq.~(\ref{eq:observable_fin_kw})---we have to perform three further steps. 
First, we convert from a correlation function of electric fields to the analogous expression for vector potentials by means of their interrelation ${\mvec{E}^{(+)}(\omega)} = i \, \alpha \, \omega \, \Acl[]{(+)}{\omega}$ in frequency space.
Next, we Fourier transform the remaining real-space dependencies.
While this is straightforward for the transverse  directions inside the source plane ($J(\mvec{x}^{\bot}_1,\mvec{x}^{\bot}_2) \rightarrow \tilde{J}(\mvec{k}^{\bot}_1,\mvec{k}^{\bot}_2)$), all information beyond $z_i = z_0$ has to be obtained from longitudinal propagation of the correlation function first.
We achieve this by means of free-space propagation, which corresponds to the simplemost case of a virtually empty beamline and an ideally transparent sample \footnote{It should be noted that the propagation of light through the sample may have significant impact on the field distribution that ultimately drives XOWM. Notable effects include beam extinction through absorption or Bragg diffraction---for instance. In the present case, however, we can neglect either, as the diamond sample is only weakly absorbing and rocked away from its Bragg-condition \cite{2012Glover-SFG}.}.
The action of an additional monochromator %, which couples spatial and spectral degrees of freedom, 
can be approximated in the spectral domain (cf. also Ref.~\cite{2017Gorobtsov-SingerCoh}) using the transmission function $T(\omega) = \text{exp} [-(\omega - \omega_m)^2 (2\Omega_m)^{-2} ]$.
In a third step, we generalize the scalar correlator in order to account for the x-rays' polarization.
For planar undulators as a source, the radiation generated on axis is linearily polarized ($\polvec{X}$) with deviations being negligible as long as the divergence of the beam is small ($\abs{\smash{\mvec{k}^{\bot}_X}}/\abs{\mvec{k}_X} \ll 1$).
Implementing all of the above steps results in the correlation function:
\begin{align}
  &({\ZH}^{(1)}_{\textsc{x\_in}}(\mvec{k}_1,\omega_1,\mvec{k}_2,\omega_2))_{\sigma\rho}
  =
  Z_0 \, (\polvec{X})_\sigma (\polvec{X})_\rho \, 
  \expo{ (k_1^{\shortparallel} - k_2^{\shortparallel}) z_0}  \nonumber \\
  &\delta(k_1^{\shortparallel} - \kappa_1) \delta(k_{2}^{\shortparallel} - \kappa_2)
  \tilde{J}(\mvec{k}^{\bot}_1,\mvec{k}^{\bot}_2) \, T(\omega_1) T(\omega_2) \, W(\omega_1,\omega_2)
  \label{eq:GSM-intermediate-result}
  .
\end{align}
Here, the prefactor $Z_0 = \bar{E}_{\text{pulse}} \sqrt{2^7 \pi^5} (\alpha \omega_1 \omega_2 \Omega \sigma^2)^{-1}$ can be related to the average pulse energy $\bar{E}_{\text{pulse}}$ in front of the monochromator by integration, while the parameters $\kappa_i = \sqrt{\alpha^2 \omega_i^2 - \smash{\abs{\smash{\mvec{k}^{\bot}_i}}}^2}$ reflect the vacuum dispersion of the beam.
Finally, we can simplify Eq.~(\ref{eq:GSM-intermediate-result}) slightly further by taking the paraxial approximation of the phase term
\begin{align}
  \expo{ (k_1^{\shortparallel} - k_2^{\shortparallel}) z_0}
  &\approx
  \expo{ (\omega_1 - \omega_2) \alpha z_0} \,
  \text{exp}{\left[ i \,  \Big( \frac{\smash{\abs{\smash{\mvec{k}^{\bot}_2}}}^2}{2 \alpha \omega_2} - \frac{\smash{\abs{\smash{\mvec{k}^{\bot}_1}}}^2}{2 \alpha \omega_1} \Big) z_0 \right]}
  \label{eq:GSM-paraxial}
  .
\end{align}
Again, this is well justified by the small beam divergence of typical undulator radiation.
Upon re-inserting this into $({\ZH}^{(1)}_{\textsc{x\_in}})_{\sigma\rho}$, we find two phase terms now - each involving the frequency difference $\omega_1 - \omega_2$. Both are related to a choice of reference frame---specifically, the spatial ($z_0$) or temporal ($t_0$) distance of the radiation's source from the respective origin. For simplicity, we center the spatial reference frame on the material sample, which fixes the upstream $z_0 < 0$.
Beyond this, we are free to choose the temporal origin such that $t_0 = z_0/c$, whereby the respective phase terms cancel each other \footnote{Note that the absolute choice of $t=0$ is arbitrary, but the \emph{relative} timing of external fields has to be maintained, of course.}.

%%%%%%%%%%%%%%%%%%%%%%%%%%%%%%%%%%%%%%%%%%
\subsection{Optical field}
\label{ssec:light-o}
%%%%%%%%%%%%%%%%%%%%%%%%%%%%%%%%%%%%%%%%%%
Analogous to the x-ray dependence, also the optical admixture enters Eqs.~(\ref{eq:observable_fin_xt}) and (\ref{eq:observable_fin_kw}) via correlation functions. These are $(\bar{G}_\textsc{opt}^{(1)})_{\nu\mu}$ or $(\bar{\ZH}^{(1)}_{\textsc{opt}})_{\nu\mu}$ for the externally driven case as well as $(\bar{S}_\textsc{opt}^{(1)})_{\nu\mu}$ or $(\bar{C}^{(1)}_{\textsc{opt}})_{\nu\mu}$ for any spontaneous contribution. Focusing on the (driven) case of SFG, we will concern ourselves only with modelling the first pair at this point \footnote{For a discussion of the spontaneous contribution,i.e., the field fluctuations captured by $(\bar{S}_\textsc{opt}^{(1)})_{\nu\mu}$ or $(\bar{C}^{(1)}_{\textsc{opt}})_{\nu\mu}$, we refer the reader to Ref.~\cite{2021Boemer-P09}.}. %XXX should we make this more prominent?
To this end, we start from the same expression as in the x-ray case, i.e., Eqs.~(\ref{eq:crossspec-14-0}) and (\ref{eq:crossspec-14}). In a first adaptation, we account for the superior coherence properties of an optical laser compared to an x-ray FEL by taking both the transverse coherence length $\xi$ and its temporal counterpart $\tau_c$ to be practically infinite ($\xi \rightarrow \infty$, $\tau_c \rightarrow \infty$). 
As a consequence, the correlation function factorizes into two electric fields of carrier frequency $\omega_0$ featuring slowly varying envelopes 
\footnote{In contrast to Eq.~(\ref{eq:tr_opt_drv}), this particular model corresponds only to one negative and one positive frequency part of the field. As such, it is analogous to the conventional definition of Glauber's coherence function \cite{1963Glauber-coherence}. This is nonetheless sufficient to describe the absorption of a photon during SFG and effectively corresponds to a rotating-wave approximation. For DFG---involving the stimulated emission of a photon---the opposite composition would obviously be required.}.
Spatially, the description reduces to a purely Gaussian beam in $\text{TEM}_{00}$ mode.
Recalling the fact that Glover et al. employ a stretched, and in consequence chirped, pulse \citep{2012Glover-SFG}, we augment each of the fields by a quadratic phase term $\text{exp} [\pm \text{i} \gamma_0 (t_i - t_0)^2]$.
This corresponds to a linear chirping of the pulse's instantaneous frequency.

Subsequently, we follow the same steps as before and obtain the optical field's correlation function
\begin{align}
  &(\bar{\ZH}^{(1)}_{\textsc{opt}}(\mvec{k}_1,\omega_1,\mvec{k}_2,\omega_2))_{\nu\mu}
  =
  Z_0 \, (\polvec{L})_\nu (\polvec{L})_\mu \,   \nonumber \\
  &\times \delta(k_1^{\shortparallel} - \kappa_1) \delta(k_{2}^{\shortparallel} - \kappa_2)
  \tilde{J}(\mvec{k}^{\bot}_1,\mvec{k}^{\bot}_2) \, W(\omega_1,\omega_2) \nonumber \\
  &\times \expo[-]{\gamma_{0} T^2 (\omega_2 - \omega_1) (2\omega_{0} - \omega_2 - \omega_1) / \tilde{\Omega}^2} \nonumber \\
  &\times \text{exp}{\left[ i \,  \Big( \frac{\smash{\abs{\smash{\mvec{k}^{\bot}_2}}}^2}{2 \alpha n \omega_2} - \frac{\smash{\abs{\smash{\mvec{k}^{\bot}_1}}}^2}{2 \alpha n \omega_1} \Big) z_0 \right]}
  \label{eq:optical-result}
  ,
\end{align}
where we have already implemented the paraxial approximation along the lines of Eq.~(\ref{eq:GSM-paraxial}). Most of the propagation phase is canceled by the choice of $t_0 =  z_0 n / c + \Delta t$ again. However, we account for the additional possibility of a time-delay $\Delta t$ between x-ray and optical pulse. This translates into a remaining phase-term, which is incorporated in $W(\omega_1,\omega_2)$.
Besides this, differences with respect to Eq.~(\ref{eq:GSM-intermediate-result}) comprise the Fourier transformed chirp-term, the omission of spectral filters $T(\omega)$ and two changes in the definitions of the longitudinal wavevector $\kappa_i$ and the prefactor $Z_0$. The wavevector should reflect the optical dispersion relation inside the sample and thus incorporates its index of refraction, i.e., $\kappa_i = \sqrt{\alpha^2 n^2 \omega_i^2 - \smash{\abs{\smash{\mvec{k}^{\bot}_i}}}^2}$.
For the prefactor, we use the new symbol $\tilde{\Omega}$ to denote the involved bandwidth [cf. $Z_0 = \bar{E}_{\text{pulse}} \sqrt{2^7 \pi^5} (\alpha \omega_1 \omega_2 \tilde{\Omega} \sigma^2)^{-1}$]. Physically, it describes the full bandwidth of the field---as ${\Omega}$ did previously in the x-ray case. Because of the stretching, however, the old $\Omega$ now takes on the role of an `instantaneous bandwidth' that corresponds to the overall stretched pulse-length $T = (2\Omega)^{-1}$. Instead the full bandwidth corresponds to the originally short pulse as  $\tilde{\Omega} = (2\tilde{T})^{-1} =  \sqrt{\Omega^2 + 4 T^2 \gamma_0^2}$.

%%%%%%%%%%%%%%%%%%%%%%%%%%%%%%%%%%%%%%%%%%
\section{Application}
\label{sec:sfg}
%%%%%%%%%%%%%%%%%%%%%%%%%%%%%%%%%%%%%%%%%%
%
Equipped with the modelling results from Sec.~\ref{ssec:matter} through \ref{ssec:light-o}, we can proceed to apply our theory. As indicated earlier, we focus on the case of sum frequency generation as observed by Glover et al. \cite{2012Glover-SFG}, on which we benchmark our above assumptions.
Starting from the generic expression (\ref{eq:observable_fin_kw}), we insert our results from Eqs.~(\ref{eq:transfer-K-to-pbc-dipole-app-fin}), (\ref{eq:GSM-intermediate-result}) and (\ref{eq:GSM-paraxial}) as well as (\ref{eq:optical-result}). 
In order to distinguish optical laser ($L$) from x-ray ($X$) pulse parameters, we designate them with respective subscripts. For instance, the pulse lengths read $T_L$ for the optical case and $T_X$ for x-rays.
Rendering the combined expression more `compact', we further choose adapted coordinates. Both for frequencies and momenta, these are given by center-of-mass and deviation coordinates like:
\begin{align}
  \bar{\omega}
  &=
  \frac{\omega_{1} + \omega_{2}}{2}
  &&\Delta \omega
  =
  \omega_{2} - \omega_{1}
  \label{eq:reduced-coord}
  .
\end{align}
Thus, we obtain for the observable:
\begin{widetext}
\begin{align}
  \langle \hat{O} \rangle 
  &= 
  \frac{\alpha^4}{V \, \omega_f \, (2\pi)^{13}} \, \abs{\polvec{f} \cdot \polvec{X}}^2 ~ 
  \int  d^3\bar{k}_L d^3\Delta k_L d^3\bar{k}_X d^3\Delta k_X \, \int  d\bar{\omega}_L d\Delta\omega_L d\bar{\omega}_X d\Delta\omega_X ~
  \delta(\omega_f - \bar{\omega}_X - \bar{\omega}_L) \, \delta(\Delta\omega_X + \Delta\omega_L) \,
    \nonumber \\
  &\times 
  Z_{0X} \, \delta(\bar{k}_X^{\shortparallel} - \bar{\kappa}_{X}^{\shortparallel}) \, \delta(\Delta k_{X}^{\shortparallel} + \Delta \kappa_{X}^{\shortparallel})  \,
  e^{-(\Delta\mvec{k}_X^{\bot} \sigma_X)^2 /2} ~
  e^{-(\bar{\mvec{k}}_X^{\bot} \delta_X)^2 /2} ~
  \expo{\frac{z_{0X} }{\alpha \, \bar{\omega}_X} \bar{\mvec{k}}_X^\bot \cdot \Delta\mvec{k}_X^{\bot}} ~
  e^{-\Delta\omega_X^2 (T_X^2 + (2\Omega_m)^{-2})/2} ~
  e^{-(\bar{\omega}_X - \omega_{0X})^2 \frac{\Omega_X^2 + \Omega_m^2}{2 \Omega_X^2 \Omega_m^2}} \, \nonumber \\
  &\times
  Z_{0L} \, \delta(\bar{k}_L^{\shortparallel} - \bar{\kappa}_{L}^{\shortparallel}) \, \delta(\Delta k_{L}^{\shortparallel} + \Delta \kappa_{L}^{\shortparallel}) \, 
  e^{-(\Delta\mvec{k}_{L}^{\bot})^2 \sigma_L^2 /2} ~
  e^{-(\bar{\mvec{k}}_{L}^{\bot})^2 \delta_L^2 /2} ~
  \expo{\frac{z_{0L} }{\alpha \, \abs{\bar{\omega}_L} \, n} \bar{\mvec{k}}_L^\bot \cdot \Delta\mvec{k}_L^{\bot}} ~  
  e^{-\Delta\omega_L^2 \tilde{T}_L^2 / 2} ~ \expo{\Delta t \, \Delta \omega_L} ~
  e^{-(\bar{\omega}_L - \omega_{0L})^2 / 2\tilde{\Omega}_L^2} \,
  \nonumber \\
  &\times
   \expo[-]{2 \gamma_{0} T_L^2 \Delta \omega_L (\omega_{0L} - \bar{\omega}_L) / \tilde{\Omega}_L^2} ~
  l_w^2 \, e^{-(k^{\shortparallel}_f -\bar{k}_{L}^{\shortparallel}     - \bar{k}_{X}^{\shortparallel} - G^{\shortparallel})^2 \sigma_w^2} \,
  e^{-(\Delta k_{L}^{\shortparallel} + \Delta k_{X}^{\shortparallel} )^2 \sigma_w^2 /4} \, 
  (2\pi)^4 \, \delta^2(\mvec{k}^{\bot}_f - \bar{\mvec{k}}^{\bot}_X - \bar{\mvec{k}}^{\bot}_L - \mvec{G}^{\bot}) \, \delta^2(\Delta\mvec{k}^{\bot}_X + \Delta\mvec{k}^{\bot}_L)  \nonumber\\
  &\times 
  \frac{1}{V_\diamond^2}
  \Big( \polvec{L} \cdot \mvec{K}_{GS\diamond}(0, \mvec{G}, \bar{\omega}_L + \Delta \omega_L /2) \Big)
  \Big( \polvec{L} \cdot \mvec{K}_{GS\diamond}(0, \mvec{G}, \bar{\omega}_L - \Delta \omega_L /2) \Big)^* 
  .
  \label{eq:obs-sdf2g-app3}
\end{align}
\end{widetext}
In writing the above, we have assumed the material system to be prepared initially in its ground state only, thus reducing the sum over $I$ to $I = GS$.
Moreover, we have approximated the crystal's shape function first by assuming a cuboid sample with factorizing $\tilde{w}(\mvec{k}) = \tilde{w}(\mvec{k}^\bot) \cdot \tilde{w}(k^{\shortparallel})$ and second by taking the transverse momentum components (parallel to the crystal surface) to be perfectly conserved, while the longitudinal constraint was approximated by a Gaussian \footnote{Notably, our approximation of the shape function $\tilde{w}(\mvec{k})$ also implies neglecting any tilt of the sample's surface normal with respect to the incident beams' direction. For the (111)-reflection measured in a (100)-cut sample, this is a decent approximation, as the deviation merely amounts to $\approx 8 \text{~degrees}$.}:
\begin{align}
  \tilde{w}(\mvec{k}^{\bot})
  &\approx
  (2\pi)^2 ~ \delta^2(\mvec{k}^{\bot}) \\
  \tilde{w}({k}^{\shortparallel}) 
  &\approx
  l_w ~ e^{-({k}^{\shortparallel})^2 \sigma_w^2 /2}
  && \text{with \hspace{2em}} 
  \sigma_w 
  =
  {l_w}/{\sqrt{2\pi}}
  .
\end{align}
For further simplification of Eq.~(\ref{eq:obs-sdf2g-app3}), we shall adopt a set of minor approximations:
The first concerns the factors $Z_{0X}$ and $Z_{0L}$, both of which contain fractions $\sim 1/(\bar{\omega} \pm \Delta \omega)$ and thus depend weakly on the respective integrations. We neglect these dependencies and choose the appropriate central frequencies $\omega_{0X}$ or $\omega_{0L}$ instead.
Next, we observe that the nonlinear response function varies slowly with frequency as long as the laser drives the system far-off any optical resonance (cf. Fig.~\ref{fig:comp-K-vs-eps}). Thus, within in the narrow bandwidth of the laser, we may similarily approximate its value by considering only the central frequency $\omega_{0L}$, i.e., \mbox{$\mvec{K}_{GS\diamond}(0, \mvec{G}, \bar{\omega}_L \pm \Delta \omega_L) \approx  \mvec{K}_{GS\diamond}(0, \mvec{G}, {\omega}_{0L})$}.
Regarding the x-ray beam, we can safely assume the spectral pass-width of monochromators to be smaller than the original SASE bandwidth and simplify \mbox{$({\Omega_X^2 + \Omega_m^2})({\Omega_X \Omega_m})^{-2} \approx \Omega_m^{-2}$}.
Finally, we can inspect the curvature phase term $\frac{z_{0X} }{\alpha \, \bar{\omega}_X} \bar{\mvec{k}}_X^\bot \cdot \Delta\mvec{k}_X^{\bot} \lessapprox  1 \ll 2\pi$ and find it negligible with modest error \footnote{Note with regard to the phase term that it may be recast as ${z_{0X} \sigma_X \delta_X}{z_{eff}^{-1}} \bar{\mvec{k}}_X^\bot \cdot \Delta\mvec{k}_L^{\bot}$ employing a momentum $\delta$-function and the definition $z_{eff} = \alpha \, \bar{\omega}_{X} \sigma_X \, \delta_X$ taken from Singer \cite{2012Singer-thesis}. Subsequently, the momenta may be estimated based on their constraining exponentials in Eq.~(\ref{eq:obs-sdf2g-app3}). Further assuming that the laser beamsize was chosen to be larger than the FEL footprint on sample results in $\frac{z_{0X}}{z_{eff}\Delta(z=0)} \lessapprox 1$---cf. again Ref.~\cite{2012Singer-thesis}.}.

Carrying out all integrations contained in Eq.~(\ref{eq:obs-sdf2g-app3}) and converting into a double differential scattering probability as per Eq.~(\ref{eq:double_diff_prob}), we arrive at:
%
%%
%\begin{align}
%  \frac{d^2P}{d\omega_f \, d\Omega_f} 
%  &\approx 
%  \Big( \frac{d\sigma}{d\Omega_f} \Big)_{\text{Th}} ~
%  A ~
%  \frac{1}{V_\diamond^2}\abs{\polvec{L} \cdot \mvec{K}_{GS\diamond}(0, \mvec{G}, \omega_{0\_L})}^2   \nonumber \\
%  &\times
%  e^{-(\omega_f - \omega_{0\_L} - \omega_{0\_X} - \omega)^2 /2 \Omega_m^2} \, 
%  e^{-{\omega}^2 / 2\tilde{\Omega}_L^2} \, \nonumber \\
%  &\times  
%  e^{-(\mvec{k}_f^{\bot} - \mvec{G}^{\bot})^2 \delta_X^2 /2}
%  e^{-(\Delta t + \omega B)^2  / 2 T_{tot}^2}
%  .
%  \label{eq:obs-sdf2g-app10}
%\end{align}
%%
%The result is proportional to the Thomson scattering cross section---denoted $(...)_{\text{Th}}$---as could be expected for a non-resonant scattering phenomenon. For brevity, we have introduced further abbreviations as follows:
%%
%\begin{align}
%  A 
%  &= 
%  \frac{4\pi \, l_w \, \omega_f \, \bar{E}_{\text{pulse}\_X} \, \bar{E}_{\text{pulse}\_L} \, \delta_X^2 \, }{(n-1) \sqrt{\pi} \,\Omega_X \, \tilde{\Omega}_L \, \omega_{X\_0}^2 \, \omega_{L\_0}^2 \, \Sigma_L^2 \, T_{tot}} \\
%  %
%  B
%  &=
%  \frac{2 \gamma_{0} T_L^2}{\tilde{\Omega}_L^2} \\
%  %
%  \omega 
%  &=
%  \frac{c k^{\shortparallel}_f - \omega_f - c G^{\shortparallel}}{(n-1)}  - \omega_{0\_L} \\
%  %
%  T_{tot} 
%  &= 
%  \sqrt{T_X^2 + (2\Omega_m)^{-2} + \tilde{T}_L^2  + [{\alpha (n-1) \sigma_w]^2/{2}}}
%  .
%\end{align}
%%  
\begin{widetext}
\begin{align}
  \frac{d^2P_\text{SFG}}{d\omega_f \, d\Omega_f} 
  &\approx 
  \Big( \frac{d\sigma}{d\Omega_f} \Big)_{\text{Th}} ~
  \frac{4\pi \, l_w \, \omega_f \, \bar{E}_{\text{pulse}\_X} \, \bar{E}_{\text{pulse}\_L} \, \delta_X^2 \, }{(n-1) \sqrt{\pi} \,\Omega_X \, \tilde{\Omega}_L \, \omega_{0X}^2 \, \omega_{0L}^2 \, \Sigma_L^2 \, T_{tot}} ~
  \frac{1}{V_\diamond^2}\abs{\polvec{L} \cdot \mvec{K}_{GS\diamond}(0, \mvec{G}, \omega_{0L})}^2   \nonumber \\
  &\times
  e^{-(\mvec{k}_f^{\bot} - \mvec{G}^{\bot})^2 \delta_X^2 /2}
  e^{-(\omega_f - \omega_{0L} - \omega_{0X} - \omega)^2 /2 \Omega_m^2} \, 
  e^{-[{\omega} + \Delta t C]^2 D / 2 \tilde{\Omega}_L^2} \, e^{-\Delta t^2  / 2 T_{tot}^2}
  .
  \label{eq:obs-sdf2g-app10}
\end{align}
\end{widetext}
First of all, we note that the result is proportional to the Thomson scattering cross section---here denoted by $(...)_{\text{Th}}$. This was to be expected for a non-resonant scattering phenomenon. In addition to the Thomson-symbol, we have introduced further abbreviations in Eq.~(\ref{eq:obs-sdf2g-app10}) as follows:
\begin{align}
  &\omega 
  =
  \frac{c k^{\shortparallel}_f - \omega_f - c G^{\shortparallel}}{(n-1)}  - \omega_{0\_L} , \\
  &C
  =
  {2 \gamma_0 T^2_L}/{T_{tot}^2}, ~~~~~~
  D
  =
  {T_{tot}^2}/({T_{tot}^2 - T_L^2 + \tilde{T}_L^2}), \\
  &T_{tot} 
  = 
  \sqrt{T_X^2 + (2\Omega_m)^{-2} + {T}_L^2  + [\alpha (n-1) \sigma_w]^2/{2}}
  \label{eq:obs-aux}
  .
\end{align}
We recall that $T_X$ and $T_L$ mark the actual lengths of x-ray and optical pulses, whereas $\tilde{T}_L$ refers to the optical pulse length prior to stretching.
Finally, we have also introduced a new symbol to mark the optical beam's size, viz. $\Sigma_L$. This refers to the laser beam's cross section at the sample position, while our original description made reference to the beam size at its waist $\sigma_L$. 

\begin{widetext}

\begin{figure}[!ht]
   \centering
   \subfloat[][]{\includegraphics[width=.24\textwidth]{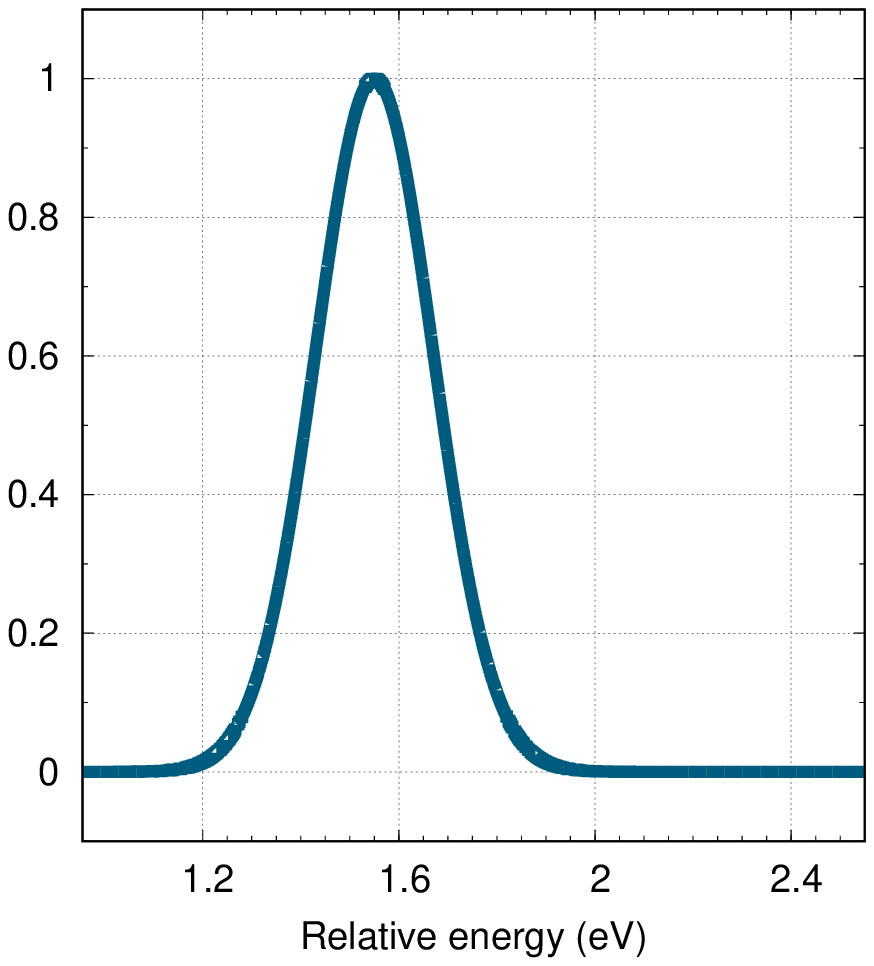}}
   \subfloat[][]{\includegraphics[width=.24\textwidth]{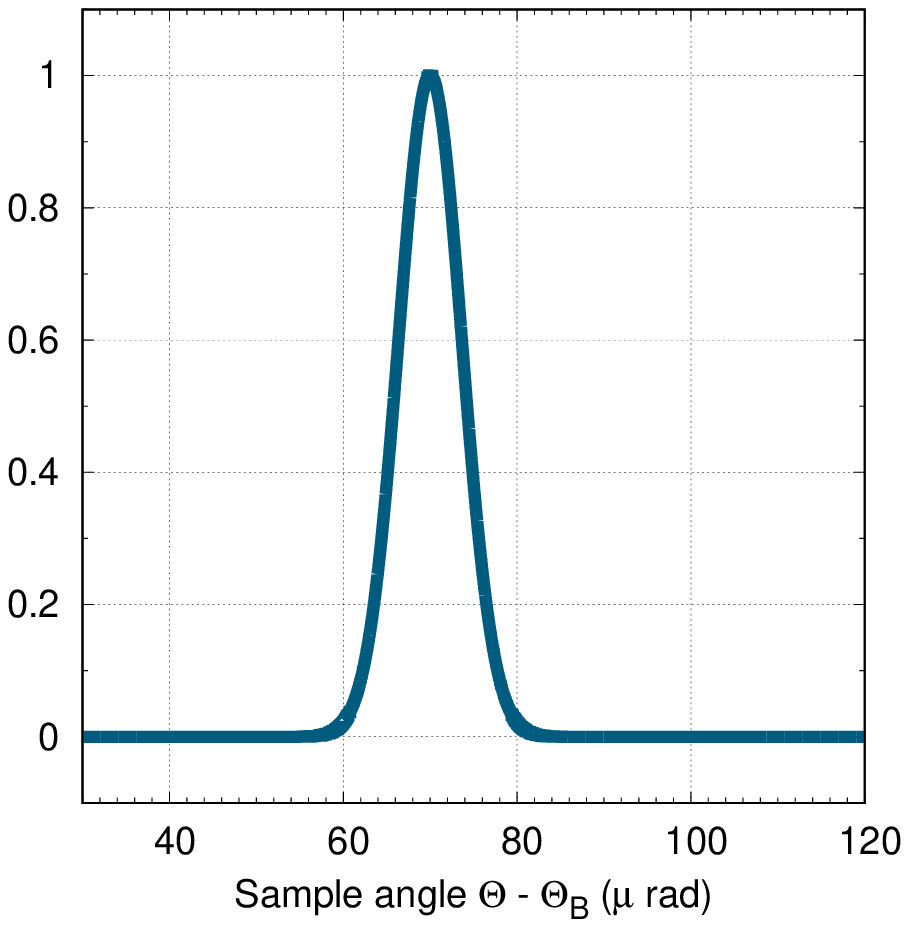}}
   \subfloat[][]{\includegraphics[width=.24\textwidth]{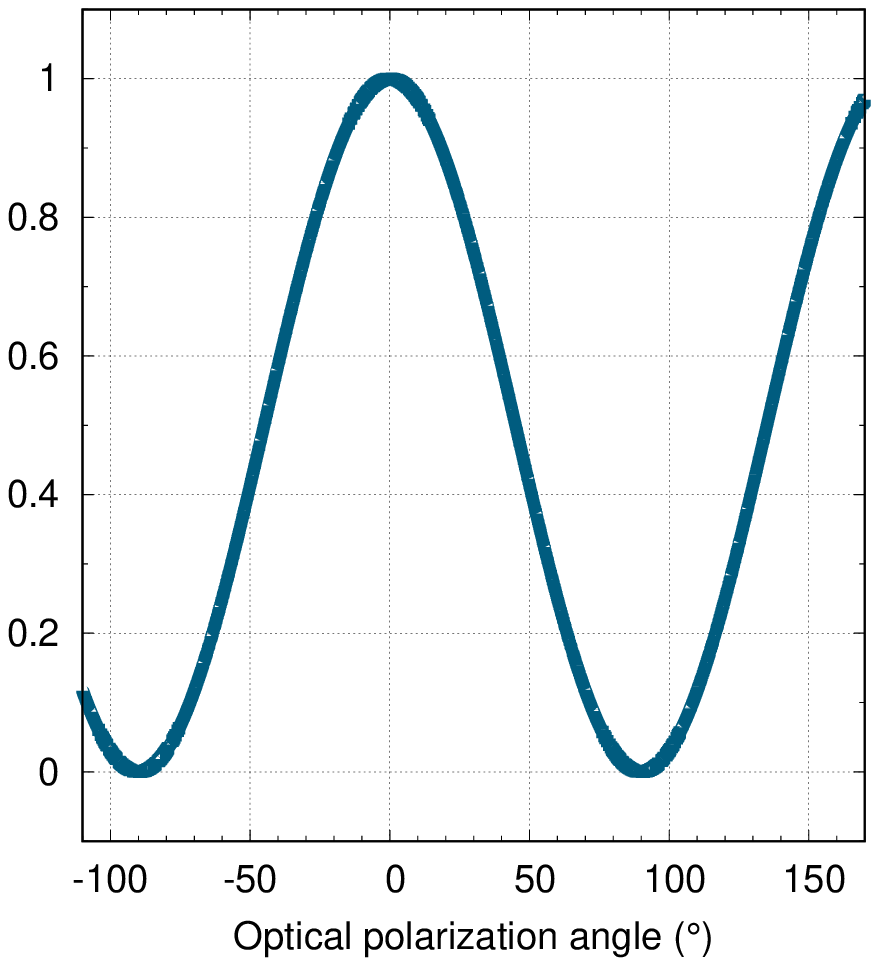}}
   \subfloat[][]{\includegraphics[width=.24\textwidth]{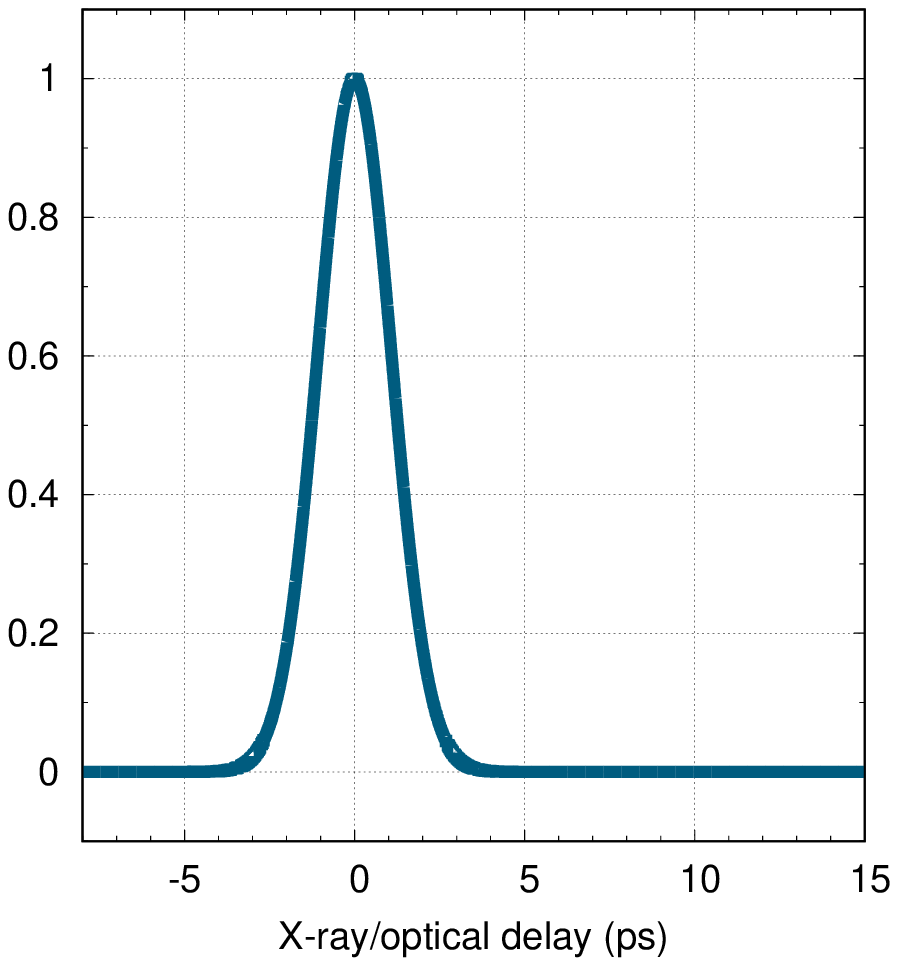}}
   \caption{Dependencies of the SFG signal (Eq.~\ref{eq:obs-sdf2g-app10}) simulated for comparison with the parameter scans of Ref.~\cite{2012Glover-SFG}. (a) Spectral dependence of SFG as measured relative to the fundamental ($\hbar \omega_{0X} = 8~\text{keV}$ with an analyzer of $0.3~\text{eV}$ passwidth (FWHM). (b) Angular behavior of SFG signal upon rotation of the sample (`rocking scan') given fixed analyzer acceptance of $17 ~\mu\text{rad}$ (FWHM). (c) Variation of SFG yield with rotation of optical polarization relative to the scattering plane ($0^\circ$ : in-plane). (d) Dependence of the SFG yield upon the temporal delay of x-ray and optical pulse. All plots are normalized for simplicity.}
   \label{fig:Glover-fake}
\end{figure}
\end{widetext}

In the following, we shall analyze our result in more detail and compare its dependencies to the observations of Glover et al.~\cite{2012Glover-SFG}. Accordingly, we set all parameters to their experimental values (see App.~\ref{sec:params} for reference). For each of the parameter scans of the original experiment (Fig.~1 in Ref.~\cite{2012Glover-SFG}), we can produce an equivalent simulation based on Eq.~(\ref{eq:obs-sdf2g-app10}). The results are shown in Fig.~\ref{fig:Glover-fake} (a) - (d) and will be discussed step-by-step hereafter.

\newpage

\paragraph{Energy dependence}
In order to separate the XOWM signal from regular elastic scattering, Glover and coworkers employ a crystal analyzer as an energy filter. Modelling their setup, we can include an additional window function $T_a(\omega_f) = e^{-(\omega_f - \omega_a)^2 /2 \Omega_a^2}$ analogous to our treatment of the x-ray monochromator. Integrating over all transmitted photon energies ($\omega_f$) eliminates the narrow exponential $e^{-[{\omega} + \Delta t C]^2 D / 2 \tilde{\Omega}_L^2}$ from Eq.~(\ref{eq:obs-sdf2g-app10}) in favor of fixing $\omega \approx 0$. Here, we disregard any possible energy shift resulting from the chirp term $\Delta t C$. This could contribute within the bandwidth of the laser ($\sim 68 ~\text{meV}$ (FWHM)) at most. At the appropriate phase-matching condition, we can now find the spectral dependence shown in Fig.~\ref{fig:Glover-fake} (a) by tuning the analyzer's central energy $\omega_a$ relative to the FEL's fundamental $\omega_{0X}$---assuming its experimental value of $\sim 8 \text{ keV}$. The resulting peak corresponds to the SFG signal that is up-shifted by $\sim \hbar \omega_{0L} = 1.55 \text{ eV}$. In the present case, the feature's width is determined by the acceptance of the analyzer ($0.3 \text{ eV}$ FWHM \cite{2012Glover-SFG}). While the SFG feature agrees very well with the experimental observation, it should be noted that our simulation does not account for the Lorentzian background that was likewise measured in Ref.~\cite{2012Glover-SFG}. However, this could be added incoherently from a dedicated x-ray-optics simulation (e.g., using the frameworks cited earlier \cite{2013Chubar-SRWWavefront,2014Klementiev-XRT,2016Samoylova-WPG}). 

\paragraph{Angular dependence}
Subsequently, we can inspect the angular characteristics of the SFG signal. This is set largely by the exponential $e^{-(\mvec{k}_f^{\bot} - \mvec{G}^{\bot})^2 \delta_X^2 /2}$ with a width $\sim 1/\delta_X$ stemming from the incident x-ray beam's divergence. Consequently, the SFG signal is emitted  as a confined beam (spread of $\sim 1.7 ~\mu\text{rad}$ \cite{2012Glover-SFG}), which precesses once the sample is being rotated. 
%This rocking-behavior may likewise be assessed from said exponential by following its peak.
%
%\begin{align}
%  0
%  &\approx
%  {k}_f^{y} - {G}^{y}
%  =
%  \alpha \omega_f \, \text{sin}(\theta_f) - \abs{\mvec{G}} \, \text{cos}(\Omega).
%\end{align}
%
Following the exponential's peak (i.e., the condition $\mvec{k}_f^{\bot} - \mvec{G}^{\bot} = 0$), we find the approximate relation
\begin{align}
  \theta_f 
  &\approx
  2 \, \theta - \frac{2 \, \alpha \, (n-1) \, \omega_{0L} \, \text{cos}(\theta)}{\abs{\mvec{G}} }
  .
\end{align}
Thus, for the sample set to $\theta$---denoting the angle among its reflecting lattice planes and the incident x-rays---the SFG signal is emitted under a scattering angle $\theta_f$. Notably, this beam precesses by twice the rocking angle of the sample ($\theta\text{-}2\theta$-behavior), which agrees with the findings of Ref.~\cite{2012Glover-SFG}. 
In order to reproduce their rocking curve, we have to convolute the traverse of the narrow SFG beam with the angular acceptance of their analyzer ($17 ~\mu\text{rad}$ FWHM) resulting in Fig.~\ref{fig:Glover-fake} (b) for our slightly simplified phase-matching condition. As a point of reference in this figure, we take the Bragg-condition for $8 ~\text{keV}$ photons reflecting off the (111)-lattice planes, i.e., the nominal Bragg angle of $\theta_B = 22.104 ~\text{degrees}$.

\paragraph{Polarization dependence}
The dependence of the SFG signal on the polarization direction of the visible light is immediately obvious from Eq.~(\ref{eq:obs-sdf2g-app10}). The scalar product of polarization vector and electronic response function implies a $\text{cos}^2$-variation as plotted in  Fig.~\ref{fig:Glover-fake} (c) and reproduces the behavior observed in Ref.~\cite{2012Glover-SFG}.

\paragraph{Delay dependence}
For the final scan, we can analyze the dependence of the signal on the mutual delay of x-ray and optical pulses based on the temporal exponential $e^{-\Delta t^2  / 2 T_{tot}^2}$ in Eq.~(\ref{eq:obs-sdf2g-app10}). This allows for the SFG signal to be created as long as the mutual delay is on the order of $T_{tot}$ (see Eq.~\ref{eq:obs-aux}), which marks the cross-correlation time of (monochromatized) x-ray and optical pulses {in vacuo} plus propagation delay. The latter results from the different speeds of light experienced by the pulses inside the sample and effectively prolongs the overlap window. Assembling $T_{tot}$ from experimental parameters and plotting the respective exponential, we obtain Fig.~\ref{fig:Glover-fake} (d) showing a full-width at half maximum of $2.53 ~\text{ps}$. This is in excellent agreement with the experimentally determined $2.5 ~\text{ps}$ (FWHM) \cite{2012Glover-SFG}. Beyond this, slight shifts of the maximum's position may be introduced as a result of the laser's chirp---encoded via the term $e^{-[{\omega} + \Delta t C]^2 D / 2 \tilde{\Omega}_L^2}$. While we have ignored this for the simplemost case, it may become relevant, if the energy acceptance on the detection side is significantly more narrow.

In addition to the specific dependencies, we also want to remark on the overall signal strength. To this end, we integrate over the spectral and angular acceptance of the analyzer setup as indicated above. Subsequently, we fix the remaining prefactor of Eq.~(\ref{eq:obs-sdf2g-app10}) in accordance with the experimental parameters (see listing in App.~\ref{sec:params}). %This requires further knowledge of the optical beam's footprint at the sample position, which amounted to $\Sigma_L^2 \approx 0.09 \text{ mm}^2$ according to the authors.
%bear in mind that the FWHM value was 0.5*1.0 mm**2 but needs to be converted to sigma.
For a $500 ~\mu\text{m}$ thick diamond sample and wavemixing near the (111)-reflection, %implying a Bragg-angle of \mbox{$\theta_B = 22.104$} degrees. With all of the above in place, 
we would predict an average yield of $Y = 0.6$ signal photons per FEL shot. This falls short of the findings by Glover and coworkers by two orders of magnitude. Besides some uncertainties in the experimental parameters, we consider most of this discrepancy to result from our approximate treatment of the material's response function $\mvec{K}_{GS}$. As this has been evaluated at DFT-level only, it suffers from the aforementioned shortcomings of bare DFT regarding excited state properties and response functions (cf. Sec.~\ref{ssec:matter}). Its low-energy spectral tail, which is relevant for the presented case ($\omega_L \ll E_{\text{bandgap}}$), could be off by an order of magnitude in particular. As any error in $\mvec{K}_{GS}$ enters quadratically into Eq.~(\ref{eq:obs-sdf2g-app10}), this would already account for the discrepancy.
In the future, this can be improved using higher-level simulations on the material part without changing the structure of the presented expressions. This flexibility, in turn, renders our approach suitable for comprehensive studies of SFG and related XOWM processes.

Going beyond previously observed features, we want to comment on a noteworthy aspect of Eq.~(\ref{eq:obs-sdf2g-app10}) in closing. Namely, that the SFG-yield depends bi-linearily on the x-ray and optical pulse energies normalized to a \emph{combined} time $T_{tot}$ and cross-sectional area---here $\Sigma_L^2$. This implies favorable scaling of the signal if \emph{both} pulses can be made as short and intense as possible. 
In the present case, the relevant time-scale $T_{tot}$ is set by the stretched optical pulse at $\sim 2 ~\text{ps}$, while the x-ray pulse length is comparably irrelevant at $\sim 80 ~\text{fs}$. As such, there is headroom left to conduct even higher-yielding SFG experiments at FEL-sources, if the optical pulse is adequatly shortened.
In contrast, no viable SFG experiments could be conceived at present storage-ring based x-ray sources. Although these could provide similar photon flux on average, the energy per pulse is lower and typically spread over a longer time ($\sim 100 ~\text{ps}$ \cite{2011Schroeder-Petra_3_bunch_length}). The latter alone makes for an intrinsic disadvantage of three orders of magnitude compared to the FEL-pulse or two orders of magnitude compared to the present case's $T_{tot}$. Besides this, the lower pulse energy could only be compensated for, if x-ray and optical pulses were mixed at a correspondingly higher repetition rate. While storage-ring based sources can deliver pulses at frequencies beyond $\sim 1~\text{MHz}$ \cite{2017Bieler-Petra_3_operation}, the same would prove challenging for a laser system to provide at $\sim 1~\text{mJ}$-level and thermally problematic for a sample to withstand.
Thus, laser driven x-ray optical wavemixing processes such as DFG and SFG appear to be ill-suited for storage-ring based x-ray sources and instead call for x-ray FELs in combination with synchronized optical laser systems to be successfully implemented.

\section{Conclusion}
Starting from first principles, viz. nonrelativistic QED, we have derived a description of XOWM in terms of a nonlinear x-ray scattering perspective. The framework allows us to treat x-ray optical SFG and DFG as well as XPDC on the same footing and reach quantitative predictions on their yields. 
We have outlined routes to model all necessary correlation functions, upon which our expression rests, and combined them to benchmark the approach on prior experimental results. Studying SFG as of Ref.~\cite{2012Glover-SFG}, we have found our description to accurately reproduce the dependence of the nonlinear scattering probability in various parameter scans. Likewise, however, we have discovered shortcomings reproducing the overall SFG-yield. Attributing this to the inaccuracies of our DFT-based model for the material's response function, we emphasize the need for higher-level simulations. 
In particular, we stress the requirement of transverse current-responses---as could be accessible via TD-CDFT \cite{2006Vignale-TD-CDFT}, for instance.
Conversely, our framework may also be used to invert experimental data and thus gain access to high-resolution data on current-density-density response functions. These can, in turn, be used to complement theoretical developments---especially for highly correlated materials studied through the lens of XOWM.

\section{Acknowledgements}
This work was supported through the Cluster of Excellence, CUI: Advanced Imaging of Matter of the ``Deutsche Forschungsgemeinschaft (DFG)---EXC 2056---project ID11390715994''. D. Krebs is part of the Max Planck School of Photonics supported by BMBF, Max Planck Society, and Fraunhofer Society.

%%%%%%%%%%%%%%%%%%%%%%%%%%%%%%%%%%%%%%%%%%
%%%%%%%%%%%%%%%%%%%%%%%%%%%%%%%%%%%%%%%%%%
%%%%%%%%%%%%%%%%%%%%%%%%%%%%%%%%%%%%%%%%%%

\appendix
%%%%%%%%%%%%%%%%%%%%%%%%%%%%%%%%%%%%%%%%%%
\section{Derivation IXS}
\label{sec:ixs}
%%%%%%%%%%%%%%%%%%%%%%%%%%%%%%%%%%%%%%%%%%
In this section, we outline the derivation of Eq.~(\ref{eq:observable_ixs_limit}) starting from Eq.~(\ref{eq:observable_reminder_U}):
\begin{align}
  \langle \hat{O}_{\mvec{k}_f,\lambda_f} \rangle (t)
  &=
  \tr{\hat{U}(t,t_0)\hat{\rho}(t_0)\hat{U}(t_0,t)\hat{O}_{\mvec{k}_f,\lambda_f}}
  .  
  \label{eq:observable_reminder_U2}
\end{align}
As outlined before, we begin by splitting the full propagators $\hat{U}(t,t_0) = \hat{U}_\textsc{0}(t,t_0) \, \hat{U}_\textsc{int}(t,t_0)$.
Here, $\hat{U}_\textsc{int}(t,t_0)$ resolves the interaction picture time dependence associated with the interaction Hamiltonian (\ref{eq:H_int})---taken solely for the x-ray parts of $\Aqu{}{\mvec{x}}$. In contrast, $\hat{U}_\textsc{0}(t,t_0)$ pertains to the time-evolution effected by all remaining elements of Eq.~(\ref{eq:H_overall}), i.e., $\hat{H}_\textsc{0} = \hat{H}_\textsc{mat}  + \hat{H}_\textsc{em} + \hat{H}_\textsc{int\_opt}$. 
Expanding $\hat{U}_\textsc{int}(t,t_0)$ to lowest non-trivial order, we obtain
\begin{align}
  \hat{U}_\textsc{int}(t,t_0)
  &\approx
  \hat{\mathbf{1}} -\text{i} \, \int_{t_0}^t \, dt_1 \, \hat{U}_\textsc{0}(t_0,t_1) \, \hat{H}_\textsc{int\_x} \, \hat{U}_\textsc{0}(t_1,t_0) 
  .  
  \label{eq:U_int_x_PT}
\end{align}
Herein, only the part of $\hat{H}_\textsc{int\_x}$ shown in Eq.~(\ref{eq:H_int_x}) can actually mediate the scattering required by our observable.
Focusing on this, we approximate the time-evolution from Eq.~(\ref{eq:observable_reminder_U2}) as:
\begin{widetext}
\begin{align}
  \langle \hat{O}_{\mvec{k}_f,\lambda_f} \rangle (t)
  \approx \, 
  &\tr{ \Big( \int_{t_0}^t dt_1 \, \hat{U}_\textsc{0}(t,t_1) \, \hat{H}_\textsc{int\_x} \, \hat{U}_\textsc{0}(t_1,t_0)  \Big)
  \hat{\rho}(t_0)
  \Big( \int_{t_0}^t dt_1^\prime \, \hat{U}_\textsc{0}(t,t_1^\prime) \, \hat{H}_\textsc{int\_x} \, \hat{U}_\textsc{0}(t_1^\prime,t_0) \Big)^\dagger
  \hat{O}_{\mvec{k}_f,\lambda_f} } \nonumber \\
  = \,
  &\alpha^4 \!\! \int_{t_0}^t \!\! dt_1 \! \int_{t_0}^t \!\! dt_1^\prime \! \int d^3x \! \int  d^3x^\prime  \,  
  \tr[sys]{ \hat{U}_\textsc{0}(t,t_1) \, \hat{n}(\mvec{x}) \, \hat{U}_\textsc{0}(t_1,t_0)  
  \, \hat{\rho}_\textsc{sys}(t_0) \,
  \hat{U}_\textsc{0}(t_0,t_1^\prime) \, \hat{n}(\mvec{x}^\prime) \, \hat{U}_\textsc{0}(t_1^\prime,t) 
  \, \hat{\mathbf{1}}_\textsc{sys} } \nonumber \\
  \times
  &\tr[x\_in]{ \hat{U}_\textsc{0}(t,t_1) \, (\Aqu[x\_in]{}{\mvec{x}})_\rho \, \hat{U}_\textsc{0}(t_1,t_0)  
  \, \hat{\rho}_\textsc{x\_in}(t_0) \,
  \hat{U}_\textsc{0}(t_0,t_1^\prime) \,  (\Aqu[x\_in]{}{\mvec{x}^\prime})_\sigma \, \hat{U}_\textsc{0}(t_1^\prime,t) 
  \,  \hat{\mathbf{1}}_\textsc{x\_in} } \nonumber \\
  \times
  &\tr[x\_out]{ \hat{U}_\textsc{0}(t,t_1) \, (\Aqu[x\_out]{}{\mvec{x}})_\rho \, \hat{U}_\textsc{0}(t_1,t_0)  
  \, \hat{\rho}_\textsc{x\_out}(t_0) \,
  \hat{U}_\textsc{0}(t_0,t_1^\prime) \,  (\Aqu[x\_out]{}{\mvec{x}^\prime})_\sigma \, \hat{U}_\textsc{0}(t_1^\prime,t) 
  \,  \hat{\Pi}_{\mvec{k}_f, \lambda_f | \textsc{x\_out}} }
 .
  \label{eq:observable_LOPT}
\end{align}
\end{widetext}
For the second step, we have made use of the factorization of $\hat{\rho}(t_0)$ as indicated in Eq.~(\ref{eq:factorize_x_rest}), while we employ Einstein's summation convention to express the dot-products $\hat{\mvec{A}}_{\textsc{x\_in}} \cdot \hat{\mvec{A}}_{\textsc{x\_out}} = (\hat{\mvec{A}}_{\textsc{x\_in}})_\sigma \, (\hat{\mvec{A}}_{\textsc{x\_out}})_\sigma$ across the factorized traces.
Furthermore, we understand each $\hat{U}_\textsc{0}$ to be curtailed to the domain of the respective trace ($\textsc{sys}$, $\textsc{x\_in}$, $\textsc{x\_out}$) and likewise $\hat{\Pi}_{\mvec{k}_f, \lambda_f | \textsc{x\_out}}$ to be the projection operator (\ref{eq:projection_observable_2}) restricted to the modes of $\hat{\mvec{A}}_{\textsc{x\_out}}$.
All these modes are---by construction---initially unoccupied, such that we can resolve the pertaining trace expression in a particularly simple form:
\begin{align}
  \tr[x\_out]{...}
  &=
  \frac{2 \pi}{V \alpha^2 \omega_f } \, (\polvec{f}^*)_\rho \, (\polvec{f})_\sigma
  \expo[-]{\mvec{k}_f\cdot(\mvec{x} - \mvec{x}^\prime)} \, \expo{\omega_f (t_1-t_1^\prime)}
  .
  \label{eq:tr_x_out_resolved}
\end{align}
To this end, we merely employ the mode decomposition of the free field (\ref{eq:em-pw-mode-decomp}) and the time-evolution of its operators
\begin{align}
  \hat{U}_\textsc{0}(t,t_1) \, \hat{a}_{\mvec{k},\lambda}^\dagger
  &=
  \hat{a}_{\mvec{k},\lambda}^\dagger \, \expo{\omega_\mvec{k}(t_1-t)} \, \hat{U}_\textsc{0}(t,t_1) 
  .
  \label{eq:free_evo_a_rel}
\end{align}
Regarding the notation, we introduce the abbreviations $\omega_{\mvec{k}_f} \rightarrow \omega_f$ and $\polvec{\mvec{k}_f,\lambda_f} \rightarrow \polvec{f}$ for improved legibility of Eq.~(\ref{eq:tr_x_out_resolved}) and following expressions.

Unlike the outgoing field, the incoming radiation is initially in a non-trivial state.
In order to allow for its description in a sufficiently general manner, yet at the same time render the $\tr[x\_in]{...}$ expression more compact, we aim for a reformulation in terms of a Glauber-type correlation function \cite{1963Glauber-coherence}.
Hereunto, we draw on the intuitive assumption that any creation of a scattered x-ray photon must imply the annihilation of an incoming x-ray photon \footnote{It should be noted that the required balance of photon creation and annihilation is mandated by energy conservation in the long-time limit. On shorter time-scales, our prescription is equivalent to a rotating wave approximation.}.
Based on this, we can restrict the first occurrence of $\hat{\mvec{A}}_{\textsc{x\_in}}$ in Eq.~(\ref{eq:observable_LOPT})
to be its positive frequency part and the second to be its negative frequency part, respectively:
\begin{widetext}
\begin{align}
  &\tr[x\_in]{ 
  \hat{U}_\textsc{0}(t,t_1) \, (\Aqu[x\_in]{(+)}{\mvec{x}})_\rho \, \hat{U}_\textsc{0}(t_1,t_0)  
  \, \hat{\rho}_\textsc{x\_in}(t_0) \,
  \hat{U}_\textsc{0}(t_0,t_1^\prime) \,  (\Aqu[x\_in]{(-)}{\mvec{x}^\prime})_\sigma \, \hat{U}_\textsc{0}(t_1^\prime,t) 
  \,  \overbrace{\hat{U}_\textsc{0}(t,t_0) \hat{U}_\textsc{0}(t_0,t)}^{\hat{\mathbf{1}}}
  } \nonumber \\
  =
  &\tr[x\_in]{ 
  (\Aqu[x\_in]{(+)}{\mvec{x},t_1})_\rho   
  \, \hat{\rho}_\textsc{x\_in} \,
  (\Aqu[x\_in]{(-)}{\mvec{x}^\prime,t_1^\prime})_\sigma  
  } 
  =
  \tr[x\_in]{ 
  \hat{\rho}_\textsc{x\_in} \,
  (\Aqu[x\_in]{(-)}{\mvec{x}^\prime,t_1^\prime})_\sigma  
  (\Aqu[x\_in]{(+)}{\mvec{x},t_1})_\rho 
  }  
  \nonumber \\
  =
  &(G_\textsc{x\_in}^{(1)}(\mvec{x}^\prime,t_1^\prime,\mvec{x},t_1))_{\sigma\rho}
  .
  \label{eq:tr_x_in}
\end{align}
\end{widetext}
Having inserted a trivial pair of time-evolution operators in the above, we can readily identify the field operators in their Heisenberg-picture form. % \footnote{XXX Maybe, I want to comment on the fact that we omit the reference time $t_0$ here and instead understand this with reference to the scattering event centered at $t=0$}.
Upon cyclic permutation of the trace, this conforms with Glauber's definition of the $G_{\sigma\rho}^{(1)}(\mvec{x}^\prime,t_1^\prime,\mvec{x},t_1)$ correlation function---except for our use of the vector potential instead of the original's electric field.

Next, we perform a similar reformulation on the remaining expression for $\tr[sys]{...}$ from Eq.~(\ref{eq:observable_LOPT}).
Again, we insert a trivial pair of time-evolution operators, identify $\hat{n}(\mvec{x},t_1)$ (and $\hat{n}(\mvec{x}^\prime,t_1^\prime)$, respectively) as Heisenberg-picture operators and abbreviate their correlator as
\begin{align}
  \tr[sys]{\hat{n}(\mvec{x},t_1) \, \hat{\rho}_\textsc{sys} \, \hat{n}(\mvec{x}^\prime,t_1^\prime) } 
  &= 
  \langle \hat{n}(\mvec{x}^\prime,t_1^\prime) \hat{n}(\mvec{x},t_1) \rangle_\textsc{sys}
  .
  \label{eq:tr_sys}
\end{align}

Joining all three results, we find from Eq.~(\ref{eq:observable_LOPT}):
\begin{align}
  \langle \hat{O}_{\mvec{k}_f,\lambda_f} \!\rangle (t)
  &=
  \frac{2 \pi \alpha^2}{V \omega_f }  (\polvec{f})_\sigma (\polvec{f}^*)_\rho \!\!
  \int_{t_0}^t \!\!\! dt_1 \!\!\! \int_{t_0}^t \!\!\! dt_1^\prime \!\!\! \int \!\! d^3\!x \!\!\! \int \!\! d^3\!x^\prime  \nonumber \\
  &\times
  \expo{\omega_f (t_1-t_1^\prime)} \,
  \expo[-]{\mvec{k}_f (\mvec{x} - \mvec{x}^\prime)} \,
  (G_\textsc{x\_in}^{(1)}(\mvec{x}^\prime,t_1^\prime,\mvec{x},t_1))_{\sigma\rho} \, \nonumber \\
  &\times
  \langle \hat{n}(\mvec{x}^\prime,t_1^\prime) \, \hat{n}(\mvec{x},t_1) \rangle_\textsc{sys}
  %\tr[sys]{ \hat{n}(\mvec{x},t_1)  \hat{\rho}_\textsc{sys}(0) \hat{n}(\mvec{x}^\prime,t_1^\prime) }
  ,
  \label{eq:observable_ixs_notyetlimit}
\end{align}
which closely resembles the aimed for Eq.~(\ref{eq:observable_ixs_limit}) already. 
For the final step, we notice that our derivation so far assumed an initial time $t_0$ as its point of reference. We may, however, translate this reference to any other instance in time, as long as we observe the accompanying time translations (via $\hat{U}_\textsc{0}$) of the initial conditions---in particular $\hat{\rho}_\textsc{x\_in}$ and $\hat{\rho}_\textsc{sys}$.
In accordance with usual scattering conventions, we take this reference to be at $t=0$, i.e., the point in time, when the light and matter subsystems (x-ray pulse and sample) centrally overlap.
Using this new reference, effectively rids Eq.~(\ref{eq:observable_ixs_notyetlimit}) of its dependence on $t_0$ and allows us to---formally---take the limits $t_0 \rightarrow -\infty$ and $t \rightarrow \infty$. Thereby, we accomodate for the fact that current x-ray detectors would not resolve the scattering process temporally, but instead integrate over the whole scattering event. The result is finally Eq.~(\ref{eq:observable_ixs_limit}).

%%%%%%%%%%%%%%%%%%%%%%%%%%%%%%%%%%%%%%%%%%
\section{Derivation of optical admixture}
\label{sec:omix}
%%%%%%%%%%%%%%%%%%%%%%%%%%%%%%%%%%%%%%%%%%
In this section, we discuss in more detail the steps leading from Eq.~(\ref{eq:correlator}) to Eq.~(\ref{eq:tr_sys_resolve}). To begin with, we use some equivalent reformulations to recast
\begin{widetext}
\begin{align}
  \langle \hat{n}(\mvec{x}^\prime,t_1^\prime) \, \hat{n}(\mvec{x},t_1) \rangle_\textsc{sys}
  &=
  \tr[sys]{\hat{\rho}_\textsc{sys}(0) \, \hat{n}(\mvec{x}^\prime,t_1^\prime) \, \hat{n}(\mvec{x},t_1)} 
  =
  \tr[sys]{\hat{n}(\mvec{x},t_1) \, \hat{\rho}_\textsc{sys}(0) \, \hat{n}(\mvec{x}^\prime,t_1^\prime) } \nonumber \\
  &=
  \tr[sys]{   
  \hat{U}_\textsc{0}(0,t_1) \, \hat{n}(\mvec{x}) \, \hat{U}_\textsc{0}(t_1,0) \,
  \hat{\rho}_\textsc{sys}(0) \,
  \hat{U}_\textsc{0}(0,t_1^\prime) \,  \hat{n}(\mvec{x}^\prime) \, \hat{U}_\textsc{0}(t_1^\prime,0) 
  } \nonumber \\
  %=
  %\tr[sys]{ \hat{n}(\mvec{x},t) \, \hat{\rho}_\textsc{sys}(0) \, \hat{n}(\mvec{x}^\prime,t^\prime) } 
  &=
  \tr[sys]{ \hat{U}_\textsc{0}(t,t_1) \, \hat{n}(\mvec{x}) \, \hat{U}_\textsc{0}(t_1,t_0)  
  \, \hat{\rho}_\textsc{sys}(t_0) \,
  \hat{U}_\textsc{0}(t_0,t_1^\prime) \,  \hat{n}(\mvec{x}^\prime) \, \hat{U}_\textsc{0}(t_1^\prime,t) 
  }
  .
  \label{eq:correlator_ini}
\end{align}
\end{widetext}
Thereby, we return to a symmetric ordering and---more importantly---re-instate the reference times $t_0$ and $t$. The latter is done to ensure consistency amongst all initial conditions used. 
Next, we assume these initial conditions to factorize for the involved subsystems, i.e., material and optical components
\begin{align}
  \hat{\rho}_\textsc{sys}(t_0)
  &\approx
  \hat{\rho}_\textsc{mat}(t_0) \otimes \hat{\rho}_\textsc{opt}(t_0)
  .
  \label{eq:factorize_mat_opt}
\end{align}
This step resembles Eq.~(\ref{eq:factorize_x_rest}), although the approximation may be far less innocuous in this case (cf. also \footnotemark[5]).
Following the factorization, we turn towards the newly isolated $\hat{\rho}_\textsc{mat}$ in order to formalize our requirement of \emph{parametric} scattering.
We consider an (XOWM) process to be parametric, if it transfers the material system from its initial state $\ket{I}$ before the scattering event to the same final state $\ket{F} = \ket{I}$ afterwards \footnote{It should be noted that this specification is fully analogous to standard discussions of elastic x-ray scattering (see, e.g., Ref.~\ref{1992Giacovazzo-BOOK}). In fact, parametric XOWM processes may be seen as a nonlinear generalization of elastic scattering.}.
Imposing this condition onto Eq.~(\ref{eq:correlator_ini}), we---first---rewrite $\hat{\rho}_\textsc{mat}(t_0) \approx \sum_I p_I \ket{I}\bra{I}$ in a diagonal representation of energy eigenstates \footnote{It should be noted that we could have employed a generally exact diagonalization of $\hat{\rho}_\textsc{mat}$ by means of Schmidt decomposition. However, the chosen approximation in terms of eigenstates is obviously more convenient in the context of perturbation theory. At the same time, it does not impose any significant limitations, as it covers typical systems in thermal equilibrium and leaves room to discuss either realistic ensembles of thermally excited (phonon-dressed) solids or just their ground state $\ket{I} = \ket{GS}$ in a first approximation.}.
Subsequently, we reformulate the respective trace
\begin{widetext}
\begin{align}
  &\tr[sys]{ \hat{U}_\textsc{0}(t,t_1) \, \hat{n}(\mvec{x}) \, \hat{U}_\textsc{0}(t_1,t_0)  
  \, \hat{\rho}_\textsc{sys}(t_0) \,
  \hat{U}_\textsc{0}(t_0,t_1^\prime) \,  \hat{n}(\mvec{x}^\prime) \, \hat{U}_\textsc{0}(t_1^\prime,t) 
  } \nonumber \\
  \approx
  &\tr[opt]{ \tr[mat]{ \hat{U}_\textsc{0}(t,t_1) \, \hat{n}(\mvec{x}) \, \hat{U}_\textsc{0}(t_1,t_0)  
  \, \hat{\rho}_\textsc{mat}(t_0) \otimes \hat{\rho}_\textsc{opt}(t_0) \,
  \hat{U}_\textsc{0}(t_0,t_1^\prime) \,  \hat{n}(\mvec{x}^\prime) \, \hat{U}_\textsc{0}(t_1^\prime,t) 
  }} \nonumber \\
  \approx
  &\tr[opt]{ \sum_F \, \bra{F} \hat{U}_\textsc{0}(t,t_1) \, \hat{n}(\mvec{x}) \, \hat{U}_\textsc{0}(t_1,t_0)  
  \, \sum_I \, p_I \, \ket{I}\bra{I} \otimes \hat{\rho}_\textsc{opt}(t_0) \,
  \hat{U}_\textsc{0}(t_0,t_1^\prime) \,  \hat{n}(\mvec{x}^\prime) \, \hat{U}_\textsc{0}(t_1^\prime,t) 
  \ket{F} } \nonumber \\
  \Rightarrow
  & \tr[opt]{ \sum_I \, p_I \bra{I} \hat{U}_\textsc{0}(t,t_1) \, \hat{n}(\mvec{x}) \, \hat{U}_\textsc{0}(t_1,t_0)  
  \, \ket{I}\bra{I} \otimes \hat{\rho}_\textsc{opt}(t_0) \,
  \hat{U}_\textsc{0}(t_0,t_1^\prime) \,  \hat{n}(\mvec{x}^\prime) \, \hat{U}_\textsc{0}(t_1^\prime,t) 
  \ket{I} } 
    ,
  \label{eq:correlator_para}
\end{align}
\end{widetext}
with the parametric constraint being implemented in the last line.
In order to obtain an explicit, perturbative expression for XOWM, we proceed by expanding each of the two transition amplitudes $\bra{I}...\ket{I}$.
Specifically, we substitute \mbox{$\hat{U}_\textsc{0}(t,t_0) = \hat{U}_\textsc{00}(t,t_0) \hat{U}_\textsc{int\_opt}(t,t_0)$}, where $\hat{U}_\textsc{00}$ captures the time-evolution due to $\hat{H}_\textsc{00} = \hat{H}_\textsc{mat}  + \hat{H}_\textsc{em}$ and $\hat{U}_\textsc{int\_opt}$ is the propagator associated with the optical light-matter interaction. 
Analogous to the x-ray case, we assume this interaction to be sufficiently weak as to allow the approximation of $\hat{U}_\textsc{int\_opt}$ through its (truncated) Dyson series (cf. the earlier Eq.~(\ref{eq:U_int_x_PT})).
In contrast to the previous treatment, however, we impose the truncation already after the leading order in $\alpha$---for each transition amplitude. 
This is sufficient for wavemixing to occur and cuts $\hat{H}_\textsc{int\_opt}$ to the form of Eq.~(\ref{eq:H_int_opt}).
Implementing the above ideas, we find for a single amplitude:
\begin{widetext}
\begin{align}
  &\bra{I} \hat{U}_\textsc{0}(t,t_1) \, \hat{n}(\mvec{x}) \, \hat{U}_\textsc{0}(t_1,t_0) \, \ket{I} \nonumber \\
  \approx
  &\bra{I} \hat{U}_\textsc{00}(t,t_1) \big[ 
  \hat{\mathbf{1}} -\text{i} \int_{t_1}^t \,\, \negs dt_2 \, \hat{U}_\textsc{00}(t_1,t_2) \, \hat{H}_\textsc{int\_opt} \, \hat{U}_\textsc{00}(t_2,t_1)\, \big]
  \, \hat{n}(\mvec{x}) \, 
  \hat{U}_\textsc{00}(t_1,t_0) \, \big[ 
  \hat{\mathbf{1}} -\text{i} \int_{t_0}^{t_1} \, \negs d\tilde{t_2} \, \hat{U}_\textsc{00}(t_0,\tilde{t_2}) \, \hat{H}_\textsc{int\_opt} \, \hat{U}_\textsc{00}(\tilde{t_2},t_0)\, \big]
  \ket{I} \nonumber \\
  %Heisenberg
=
 &\bra{I} \hat{U}_\textsc{00}(t,0) \big[ 
  \hat{\mathbf{1}} -\text{i} \int_{t_1}^t \,\, \negs dt_2 \, \hat{H}_\textsc{int\_opt}(t_2) \big]
  \, \hat{n}(\mvec{x},t_1) \, 
  \big[ 
  \hat{\mathbf{1}} -\text{i} \int_{t_0}^{t_1} \, \negs d\tilde{t_2} \, \hat{H}_\textsc{int\_opt}(\tilde{t_2}) \big] 
  \hat{U}_\textsc{00}(0,t_0)
  \ket{I} \nonumber \\
  \approx
 &\bra{I} \hat{U}_\textsc{00}(t,0) \, (-\text{i}) \, \alpha \, \Big( 
  \int_{t_1}^t \,\, \negs dt_2 \! \int \! d^3y \, \hat{\mvec{p}}(\mvec{y},t_2) \! \cdot \! \Aqu[opt]{}{\mvec{y},t_2} \, \hat{n}(\mvec{x},t_1) 
  + 
  \hat{n}(\mvec{x},t_1) \int_{t_0}^{t_1} \, \negs d\tilde{t_2} \! \int \! d^3\tilde{y} \, \hat{\mvec{p}}(\mvec{\tilde{y}},\tilde{t_2}) \! \cdot \! \Aqu[opt]{}{\mvec{\tilde{y}},\tilde{t_2}} 
  \Big)
  \hat{U}_\textsc{00}(0,t_0)
  \ket{I} \nonumber \\ 
  = 
 &\bra{I} \hat{U}_\textsc{00}(t,0) \, (-\text{i}) \, \alpha  
  \int_{t_0}^t \,\, \negs dt_2 \! \int \! d^3y \, \hat{T} \left[(\hat{\mvec{p}}(\mvec{y},t_2))_\mu \, \hat{n}(\mvec{x},t_1)\right]
  (\Aqu[opt]{}{\mvec{y},t_2})_\mu 
  \hat{U}_\textsc{00}(0,t_0)
  \ket{I} 
  .
  \label{eq:amplitude_para}
\end{align}
Here, we have once again introduced time-dependent operators for brevity.
In contrast to the pure x-ray case, however, they are defined with respect to $\hat{H}_\textsc{00}$ now, i.e., \mbox{$\hat{n}(\mvec{x},t) = \hat{U}_\textsc{00}(0,t) \, \hat{n}(\mvec{x}) \, \hat{U}_\textsc{00}(t,0)$}.
Following the substitution of $\hat{H}_\textsc{int\_opt}$ (\ref{eq:H_int_opt}), we neglect both the trivial term and all contributions of order $\alpha^2$ and higher---as outlined before.
In the last step, we employ the usual time-ordering operator \cite{1995PeskinSchroeder-BOOK} to render Eq.~(\ref{eq:amplitude_para}) more compact.

Finally, combining our results across Eqs.~(\ref{eq:correlator_ini}), (\ref{eq:correlator_para}) and (\ref{eq:amplitude_para}), we arrive at:
%%%
\begin{align}
  \langle \hat{n}(\mvec{x}^\prime,t_1^\prime) \, \hat{n}(\mvec{x},t_1) \rangle_\textsc{sys} 
  &= \int_{t_0}^t \!\!\! dt_2 \!\!\! \int_{t_0}^t \!\!\! dt_2^\prime \!\!\! \int \!\! d^3y \!\!\! \int \!\! d^3y^\prime \,%\nonumber \\
  \alpha^2 \,   
  \tr[opt]{ 
  (\Aqu[opt]{}{\mvec{y},t_2})_\mu \, \hat{\rho}_\textsc{opt}(0) \, (\Aqu[opt]{}{\mvec{y}^\prime,t_2^\prime})_\nu
  } \nonumber \\
  &\times
  \sum_{I} \, p_I 
  \left( \PProp{I} (\mvec{y},t_2, \, \mvec{x},t_1) \right)_\mu
  \,
  \left( \PProp{I} (\mvec{y}^\prime,t_2^\prime, \, \mvec{x}^\prime,t_1^\prime) \right)_\nu^*
  \label{eq:tr_sys_resolve_t0}
\end{align}
with the correlation function $\PProp{I} (\mvec{y},t_2, \, \mvec{x},t_1)$ as defined in Eq.~(\ref{eq:PProp}).
As a last refinement, we can proceed analogous to App.~\ref{sec:ixs} again and take the limits $t_0 \rightarrow -\infty$ and $t \rightarrow \infty$, 
thus obtaining the sought after Eq.~(\ref{eq:tr_sys_resolve}).

%%%%%%%%%%%%%%%%%%%%%%%%%%%%%%%%%%%%%%%%%%
\section{Fourier relations}
\label{sec:Fourier}
%%%%%%%%%%%%%%%%%%%%%%%%%%%%%%%%%%%%%%%%%%
%

Augmenting the discussion of Sec.~\ref{sec:theory}, we list all of the Fourier relations that were employed to obtain Eq.~(\ref{eq:observable_fin_kw}):
\begin{align}
  (G_\textsc{x\_in}^{(1)}(\mvec{x}^\prime,t_1^\prime,\mvec{x},t_1))_{\sigma\rho}
  &=
  \frac{1}{(2\pi)^8} ~ \int \!\! d^3k_X  \int \!\! d^3k_X^\prime  \int \!\! d\omega_X  \int \!\! d\omega_X^\prime
  ~ \expo[-]{(\mvec{k}_X^\prime\cdot\mvec{x}^\prime - \omega_X^\prime t_1^\prime)} ~ \expo{(\mvec{k}_X\cdot\mvec{x} - \omega_X t_1)} ~
  (\ZH_\textsc{x\_in}^{(1)}\argline{\prime}{X}{\prime}{X}{}{X}{}{X})_{\sigma\rho} \\
  (\bar{G}_\textsc{opt}^{(1)}(\mvec{y}^\prime,t_2^\prime,\mvec{y},t_2))_{\nu\mu}
  &=
  \frac{1}{(2\pi)^8} ~ \int \!\! d^3k_O  \int \!\! d^3k_O^\prime  \int \!\! d\omega_O  \int \!\! d\omega_O^\prime
  ~ \expo[-]{(\mvec{k}_O^\prime\cdot\mvec{y}^\prime - \omega_X^\prime t_2^\prime)} ~ \expo{(\mvec{k}_O\cdot\mvec{y} - \omega_O t_2)} ~
  (\bar{\ZH}_\textsc{opt}^{(1)}\argline{\prime}{O}{\prime}{O}{}{O}{}{O})_{\nu\mu} \\
  (\bar{S}_\textsc{opt}^{(1)}(\mvec{y}^\prime,t_2^\prime,\mvec{y},t_2))_{\nu\mu}
  &=
  \frac{1}{(2\pi)^8} ~ \int \!\! d^3k_O  \int \!\! d^3k_O^\prime  \int \!\! d\omega_O  \int \!\! d\omega_O^\prime
  ~ \expo[-]{(\mvec{k}_O^\prime\cdot\mvec{y}^\prime - \omega_X^\prime t_2^\prime)} ~ \expo{(\mvec{k}_O\cdot\mvec{y} - \omega_O t_2)} ~
  (\bar{C}_\textsc{opt}^{(1)}\argline{\prime}{O}{\prime}{O}{}{O}{}{O} )_{\nu\mu}\\
  \PProp{I} (\mvec{y},t_2, \, \mvec{x},t_1) 
  &=
  \PProp{I} (\mvec{y}, 0, \, \mvec{x},(t_1 - t_2))
  =
  \frac{1}{(2\pi)^7} ~ \int \!\! d^3k_1  \int \!\! d^3k_2  \int \!\! d\omega
  ~ \expo{(\mvec{k}_1\cdot\mvec{y} + \mvec{k}_2\cdot\mvec{x})} ~ \expo[-]{\omega(t_1 - t_2)} ~
  \mvec{K}_I(\mvec{k}_1,\mvec{k}_2,\omega) \\
  \PProp{I\diamond} (\mvec{x}_1,0, \, \mvec{x}_2,\tau)
  &=
  \frac{1}{V_\diamond^2} ~
  \sum^{\text{1. BZ}}_{\mvec{q}} \, \sum^{\text{rec. latt.}}_{\mvec{G}_1,\mvec{G}_2} \,  \int \!\! d\omega ~
  \expo{(\mvec{q} + \mvec{G}_1)\cdot\mvec{x}_1} \, \expo{(-\mvec{q} + \mvec{G}_2)\cdot\mvec{x}_2} ~ \expo[-]{\omega \tau} ~
  \mvec{K}_{I\diamond}(\mvec{q} + \mvec{G}_1, -\mvec{q} + \mvec{G}_2,\omega)
  .
  \end{align}
In the second-last line, we have made use of the time-translation invariance of $\PProp{I}$ and in the last line, we have additionally implied that $\PProp{I\diamond}$ is lattice periodic for simultaneous translation of both spatial arguments.
%\end{widetext}

%\begin{widetext}
%%%%%%%%%%%%%%%%%%%%%%%%%%%%%%%%%%%%%%%%%%
\section{Experimental parameters}
\label{sec:params}
%%%%%%%%%%%%%%%%%%%%%%%%%%%%%%%%%%%%%%%%%%
Supplementing our discussion of Sec.~\ref{sec:sfg}, we list the relevant parameters used to characterize the experiment of Glover et al.. Drawing from Ref.~\cite{2012Glover-SFG}, we obtain for the x-ray pulse's properties:
\begin{center}
  \begin{tabular}{ | l | c | c |}
    \hline
        & experimental value & parameter in a.u. \\ \hline \hline
    photon energy & $8 ~\text{keV}$ & $\omega_{0X} = 294$ \\ \hline
    initial bandwidth & $20 ~\text{eV}$ (FWHM) & $\Omega_X = 0.312$ \\ \hline
    monochromator passwidth & $1 ~\text{eV}$ (FWHM) & $\Omega_m = 0.016$ \\ \hline
    pulse duration & $80 ~\text{fs}$ (FWHM) & $T_X = 1404$\\ \hline
    average pulse energy & $5\cdot 10^{10}$ photons & $E_{\text{pulse}X} = 1.47 \cdot 10^{13}$ \\ \hline
    beam divergence & $1.7 ~\text{$\mu$rad}$ &  $\delta_X = 6.46 \cdot 10^{5}$ \\
    \hline
  \end{tabular}
\end{center}
The optical pulse is respectively characterized by:
\begin{center}
  \begin{tabular}{ | l | c | c |}
    \hline
         & experimental value & parameter in a.u. \\ \hline \hline   
    photon energy & $1.55 ~\text{eV}$ & $\omega_{0L} = 5.7 \cdot 10^{-2}$ \\ \hline
    full bandwidth     & $68 ~\text{meV}$ (FWHM) & $\tilde{\Omega}_L = 1.06 \cdot 10^{-3}$ \\ \hline
    pulse duration (stretched) & $2 ~\text{ps}$ (FWHM) & $T_L = 3.51 \cdot 10^{5}$\\ \hline
    pulse energy  & $1 ~\text{mJ}$ &  $E_{\text{pulse}L} = 2.29 \cdot 10^{14}$\\ \hline
    beam size on sample \footnote{private communication}& $\sim 0.5 \times 1.0 ~\text{mm}^2$ (FWHM) & $\Sigma_L^2 = 3.22 \cdot 10^{13}$     \\ 
    \hline
  \end{tabular}
\end{center}
In addition, we recall that the sample was a $500 ~\text{$\mu$m}$-thick diamond furnished with a (100)-surface cut and used on a (111)-reflection in transmission geometry. The nominal Bragg angle for this reflection at $8 ~\text{keV}$ is $\theta_B = 22.104$ degrees. Finally, the analyzer used to discriminate the SFG-signal was specified to provide a spectral passwidth of $0.3 ~\text{eV}$ (FWHM)---corresponding to $\Omega_m = 4.8 \cdot 10^{-3} ~\text{a.u.}$---and an angular acceptance of $17 ~\mu\text{rad}$ (FWHM).

\end{widetext}

%\begin{align}
%  \omega_{\textsc{nlc}}
%  &=
%  \frac{n ~ \omega_{\text{in}}}{1 + (n ~ \alpha^2 \omega_{\text{in}} )(1 - \text{cos}~\vartheta)}
%  ,
%  \label{eq:intr-KB}
%\end{align}

%\include(citation}

% use bibtex
%\bibliographystyle{plainnat}
%\bibliographystyle{apsrev4-2}
\bibliography{literature_xowm}
%\begin{thebibliography}{4}

\end{document}